# Representative sets and irrelevant vertices: New tools for kernelization

Stefan Kratsch[*][†]      Magnus Wahlström[‡]

June 27, 2012

**Abstract**

The existence of a polynomial kernel for ODD CYCLE TRANSVERSAL was a notorious open problem in parameterized complexity. Recently, this was settled by the present authors (Kratsch and Wahlström, SODA 2012), with a randomized polynomial kernel for the problem, using matroid theory to encode flow questions over a set of terminals in size polynomial in the number of terminals (rather than the total graph size, which may be superpolynomially larger).

In the current work we further establish the usefulness of matroid theory to kernelization by showing applications of a result on *representative sets* due to Lovász (Combinatorial Surveys 1977) and Marx (TCS 2009). We show how representative sets can be used to give a polynomial kernel for the elusive ALMOST 2-SAT problem (where the task is to remove at most $k$ clauses to make a 2-CNF formula satisfiable), solving a major open problem in kernelization.

We further apply the representative sets tool to the problem of finding *irrelevant vertices* in graph cut problems, that is, vertices which can be made undeletable without affecting the status of the problem. This gives the first significant progress towards a polynomial kernel for the MULTIWAY CUT problem; in particular, we get a kernel of $\mathcal{O}(k^{s+1})$ vertices for MULTIWAY CUT instances with at most $s$ terminals.

Both these kernelization results have significant spin-off effects, producing the first polynomial kernels for a range of related problems.

More generally, the irrelevant vertex results have implications for covering min-cuts in graphs. For a directed graph $G = (V, E)$ and sets $S, T \subseteq V$, let $r$ be the size of a minimum $(S, T)$-vertex cut (which may intersect $S$ and $T$). We can find a set $Z \subseteq V$ of size $\mathcal{O}(|S| \cdot |T| \cdot r)$ which contains a minimum $(A, B)$-vertex cut for every $A \subseteq S, B \subseteq T$. Similarly, for an undirected graph $G = (V, E)$, a set of terminals $X \subseteq V$, and a constant $s$, we can find a set $Z \subseteq V$ of size $\mathcal{O}(|X|^{s+1})$ which contains a minimum multiway cut for any partition of $X$ into at most $s$ pairwise disjoint subsets (see the paper for a detailed description). Both results are polynomial time. We expect this to have further applications; in particular, we get direct, reduction rule-based kernelizations for all problems above, in contrast to the indirect compression-based kernel previously given for ODD CYCLE TRANSVERSAL.

All our results are randomized, with failure probabilities which can be made exponentially small in $n$, due to needing a *representation* of a matroid to apply the representative sets tool.

---

[*]Supported by the Netherlands Organization for Scientific Research (NWO), project "KERNELS".
[†]Utrecht University, Utrecht, the Netherlands. `s.kratsch@uu.nl`
[‡]Max-Planck-Institute for Informatics, Saarbrücken, Germany. `wahl@mpi-inf.mpg.de`

# 1 Introduction

Polynomial kernelization is a formalization of the notion of efficient polynomial-time preprocessing, or more generally of efficient instance simplification and data reduction. Such reduction steps are commonly applied in practice, see, e.g., the well-known CPLEX integer programming package, or many state-of-the-art SAT solvers. However, to study this theoretically, one needs a notion of the *hardness* of an instance beyond the instance size, e.g., the length of a certificate [20] or a more generic *parameter* associated with the input (cf. [12, 33]). Informally, a kernelization is a polynomial-time reduction of an input instance, with parameter value $k$, to an equivalent instance of the same problem, the *kernel*, with total output size bounded as a function of $k$; a problem has a *polynomial kernel* if the size bound is polynomial in $k$. This turns out to be a robust and interesting notion, and there is much work on both upper and lower bounds for the existence of, or best possible size of, a polynomial kernel for various problems; see [4, 13] and [3, 14, 10].

Among the problems for which the existence of polynomial kernels is still open, one can identify two major groups. The first group is centered around the ALMOST 2-SAT problem: Given a 2-CNF formula $\mathbb{F}$ and an integer $k$, can you remove at most $k$ clauses to make $\mathbb{F}$ satisfiable (or, equivalently, find an assignment under which at most $k$ clauses are not satisfied)? This is a natural, expressive problem which (at least for purposes of parameterized complexity and kernelization) captures several problems of independent interest. For one thing, it directly expresses ODD CYCLE TRANSVERSAL (OCT); the existence of a polynomial kernel for OCT was a long-standing open problem, only recently solved by the present authors [25]. Less directly, a polynomial kernel for ALMOST 2-SAT has been shown to imply the same for VERTEX COVER ABOVE MATCHING, KÖNIG VERTEX DELETION for graphs with perfect matchings, and the RHORN-BACKDOOR DELETION SET problem from practical SAT solving (cf. [11, 24]), among other problems; see [38, 16, 32]. We add to the list VERTEX COVER ABOVE LP, i.e., VERTEX COVER parameterized by the size of the LP gap. For all of these problems, no polynomial kernel was previously known.

The second group of open problems represents the class of *graph cut* problems. This is a wide class, where little is known regarding polynomial kernelization; problems for which polynomial kernelization is open include DIRECTED FEEDBACK VERTEX SET (arguably one of the biggest open problems in kernelization; see [34]), MULTIWAY CUT, and MULTICUT under various parameterizations, as well as GROUP FEEDBACK ARC/VERTEX SET, which again generalizes OCT.

In this paper, we show polynomial kernels for ALMOST 2-SAT, and for a collection of graph cut problems, including MULTIWAY CUT with a constant number of terminals and MULTICUT with a constant number of cut requests. We also show results about covering min-cuts and multiway cuts through a set of terminals using few vertices, which should be of independent interest. We make use of a lemma on *representative sets* from matroid theory, due to Lovász [26] and Marx [28]. In particular, we show how to apply the lemma in *irrelevant vertex* arguments, i.e., how to use it to find vertices in cut problems which can be made undeletable without affecting the outcome. All our results are randomized, with failure probabilities which can be made exponentially small in the input size.

**Related work.** ALMOST 2-SAT (also known as MIN 2CNF DELETION) was showed to be FPT, runtime $\mathcal{O}^*(15^k)$, by Razgon and O'Sullivan [39]; this has been improved to $\mathcal{O}^*(9^k)$ [38], $\mathcal{O}^*(4^k)$ [8], and $\mathcal{O}^*(2.6181^k)$ [32]. It has an $\mathcal{O}(\sqrt{\log n})$-approximation by Agarwal et al. [1], and no constant factor approximation under the unique games conjecture [23].

Graph cut problems have been a catalyst for the development of new techniques in parameterized complexity, including the now ubiquitous *iterative compression* technique [40, 18], the notion of *important separators* [27], and the *shadow removal* technique [29]. Our focus here is on MULTIWAY CUT($k$), first showed to be FPT by Marx [27]. The currently fastest algorithm [8], runtime $\mathcal{O}^*(2^k)$, uses an LP approach based on work of Guillemot [17]; we also use some insights of the latter.

As for polynomial kernelization of graph cut problems, in joint work with Cygan, Pilipczuk, and Pilipczuk [7] the present authors show, amongst others, that MULTICUT($k$) and DIRECTED 2-MULTIWAY CUT($k$) do not admit polynomial kernels unless the polynomial hierarchy collapses. Apart from this, and previous work [25] for ODD CYCLE TRANSVERSAL, little is known about kernels for cut or feedback problems beyond the kernelizations for FEEDBACK VERTEX SET, e.g, [44].

Matroids have seen little use as tools in parameterized algorithms (though see [30]), and only few papers address problems *on* matroids. However, recent work of Marx [28] on a parameterized matroid intersection problem also provides some results that are used in the current paper. Regarding kernelization, to our best knowledge, previous work of the present authors [25] is the first and so far only application of matroid theory, using it to encode terminal cut functions of a (large) graph into small space.

Irrelevant vertex arguments are a central part of the DISJOINT PATHS algorithm of Robertson and Seymour [41], which lies behind the celebrated FPT algorithm for testing graph minors. However, the arguments used by Robertson and Seymour to locate irrelevant vertices are very different from those used in this paper (and the resulting bounds are far from polynomial).

**Our results.** We show several applications of the representative sets lemma of Lovász [26] and Marx [28] to polynomial kernelization, producing the first polynomial kernels for a range of important problems. First, we study DIGRAPH PAIR CUT, a constrained graph cut problem designed to capture (the iterative compression form of) ALMOST 2-SAT. We show that the representative sets lemma can be used to simplify DIGRAPH PAIR CUT down to a cut problem involving a polynomial number of terminals; from here, we can get a polynomial kernel either by compression methods as in [25] or a direct kernel by the cut-covering sets given below. We get the following.

**Theorem 1.** ALMOST 2-SAT *with a bound $k$ on the solution size has a polynomial-time randomized compression into size $\tilde{\mathcal{O}}(k^6)$, with one-sided error probability $\mathcal{O}(2^{-k})$ and false positives only, and a randomized kernel with $\mathcal{O}(k^6)$ variables and failure probability $\mathcal{O}(2^{-n})$.*

As mentioned above, this gives the first polynomial kernels for a range of problems. Next, we apply the representative sets lemma to the search for irrelevant vertices for graph cut problems. This gives two sets of results. The first relates directly to polynomial kernelization.

**Theorem 2.** *The following kernelizations are possible:* MULTIWAY CUT WITH DELETABLE TERMINALS($k$), *with $\mathcal{O}(k^3)$ vertices;* $s$-MULTIWAY CUT($k$), *with $\mathcal{O}(k^{s+1})$ vertices;* $s$-MULTICUT($k$), *with $\mathcal{O}(k^{\lceil\sqrt{2s}\rceil+1})$ vertices;* GROUP FEEDBACK VERTEX SET($k$), *for a group of $s$ elements, with $\mathcal{O}(k^{2s+2})$ vertices. All results are randomized, with failure probability exponentially small in $n$.*

Finally, as a second set of irrelevant vertex results, we get interesting conclusions about covering min-cuts and multiway cuts in graphs.

**Theorem 3.** *Let $G = (V, E)$ be a digraph and let $S, T \subseteq V$. Let $r$ denote the size of a minimum $(S, T)$-vertex cut (which may intersect $S$ and $T$). There exists a set $Z \subseteq V$, $|Z| = \mathcal{O}(|S|\cdot|T|\cdot r)$, such that for any $A \subseteq S$ and $B \subseteq T$, it holds that $Z$ contains a minimum $(A, B)$-vertex cut. We can find such a set in randomized polynomial time with failure probability $\mathcal{O}(2^{-n})$.*

**Theorem 4.** *Let $G = (V, E)$ be an undirected graph and $X \subseteq V$. For any $s$, there exists a set $Z \subseteq V$, $|Z| = \mathcal{O}(|X|^{s+1})$, such that for any partition $\mathcal{X} = (X_1, \ldots, X_s)$ with pairwise disjoint subsets of $X$, it holds that $Z$ contains a minimum multiway cut of $\mathcal{X}$ (i.e., a minimum cut $C$ such that no pairs of sets $X_i, X_j$ are connected to each other in $G - C$). We can find such a set in randomized polynomial time with failure probability $\mathcal{O}(2^{-n})$.*

**Organization.** Our paper is organized as follows. Section 2 contains preliminaries, and Section 3 presents the matroid theory tools we use. Section 4 gives the first application, in the form of a polynomial kernel for ALMOST 2-SAT, and Section 5 gives irrelevant vertex-type consequences, yielding polynomial kernels for variants of MULTIWAY CUT and the cut-covering sets of Theorems 3 and 4. Sections 6 and 7 give further consequences of the ALMOST 2-SAT kernel and irrelevant vertex results, Section 8 contains delayed full proofs, and Section 9 concludes the paper.

## 2 Preliminaries

**Parameterized complexity and kernelization.** A *parameterized problem* is a language $\mathcal{Q} \subseteq \Sigma^* \times \mathbb{N}$; the second component of instances $(x, k)$ is called the parameter (cf. [12, 33]). A parameterized problem is *fixed-parameter tractable* (FPT) if there is an algorithm $A$ and a computable function $f \colon \mathbb{N} \to \mathbb{N}$ such that $A$ decides $(x, k) \in \mathcal{Q}$ in time $f(k)|x|^{\mathcal{O}(1)}$. A *kernelization* of $\mathcal{Q}$ is a polynomial-time computable mapping $K \colon \Sigma^* \times \mathbb{N} \to \Sigma^* \times \mathbb{N} \colon (x, k) \mapsto (x', k')$ such that $(x, k) \in \mathcal{Q}$ if and only if $(x', k') \in \mathcal{Q}$ and with $|x'|, k' \leq h(k)$ where $h$ is a computable function; $h$ is called the *size* of the kernel and $K$ is a *polynomial kernelization* if $h(k)$ is polynomially bounded.

All kernelization results in this paper are randomized, i.e., there is a (small) chance for the reduced instance not to be equivalent to the input. In all cases, the failure is either one-sided, with false positives only, or occurs with probability exponentially small in the input size. The former type of kernels were called coRP-kernels in previous work [25]; see [25] for a brief discussion on why they are compatible with the lower bound framework [3, 14], and [20, 14] for more on randomized compression. The latter type is easily seen to be computable in non-uniform polynomial time, compatible with the exclusion of non-uniform compression [14].

**Matroids.** A *matroid* is a pair $M = (E, \mathcal{I})$, where $E$ is the *ground set* and $\mathcal{I} \subseteq 2^E$ a collection of *independent sets*, such that: (i) $\emptyset \in \mathcal{I}$; (ii) if $I_1 \subseteq I_2$ and $I_2 \in \mathcal{I}$, then $I_1 \in \mathcal{I}$; and (iii) if $I_1, I_2 \in \mathcal{I}$ and $|I_2| > |I_1|$, then there exists some $x \in (I_2 \setminus I_1)$ such that $I_1 \cup \{x\} \in \mathcal{I}$. A set $I \subseteq E$ is *independent* if $I \in \mathcal{I}$, and *dependent* otherwise. A set $B \in \mathcal{I}$ is a *basis* of $M$ if no superset of $B$ is independent; a matroid may equivalently be defined by its set of bases. For a subset $X \subseteq E$, the *rank* $r(X)$ of $X$ is the largest cardinality of an independent set $I \subseteq X$. The rank of $M$ is $r(M) := r(E)$.

Let $A$ be a matrix over a field $\mathbb{F}$ and $E$ be the set of columns of $A$. Let $\mathcal{I}$ be the set of all sets $X \subseteq E$ of columns that are linearly independent over $\mathbb{F}$ (as vectors). Then $(E, \mathcal{I})$ defines a matroid $M$, and we say that $A$ *represents* $M$. A matroid is *representable* (over a field $\mathbb{F}$) if there is a matrix (over $\mathbb{F}$) that represents it. A matroid representable over some field is called *linear*. In this work, we will concern ourselves only with linear matroids.

**Gammoids.** Let $D = (V, A)$ be a digraph and let $S, T \subseteq V$. The set $T$ is *linked to* $S$ if there exist $|T|$ vertex-disjoint paths from $S$ to $T$; this allows paths of length zero, e.g., any set is linked to itself. Given any digraph $D = (V, A)$ with source vertices $S \subseteq V$, the sets $T \subseteq V$ which are linked to $S$ in $D$ form a matroid, a so-called *gammoid* [36] (see also [35, 42]); we refer to it as $(D, S)$. Marx [28] gave a randomized polynomial-time procedure for finding a representation of

a gammoid. The error probability can be made exponentially small in the size of the graph. By standard arguments, advice polynomial in $n$ is sufficient to derandomize this step.

**Theorem 5** ([36, 28]). *Let $D = (V, A)$ be a directed graph, and let $S \subseteq V$. The subsets $T \subseteq V$ which are linked to $S$ form the independent sets of a matroid over $V$. Furthermore, a representation of this matroid can be obtained in randomized polynomial time with one-sided error.*

Throughout the paper, $(A, B)$-cuts may intersect $A$ and $B$, unless otherwise noted. We also create *sink-only copies* of vertices; a sink-only copy of $v \in V$, in $D = (V, A)$, is a parallel copy $v'$ of $v$ which retains only the incoming edges of $v$ (in the undirected case all edges are oriented inwards). These will be used, effectively, to require two paths to a vertex $v \in X$ in a set $X$ linked to $S$. Note that adding $v'$ to the graph has no effect on any independent (linked) set not containing $v'$.

## 3 Tools from matroid theory

**Representative sets.** The notion of *representative sets* plays an essential role in the paper.

**Definition 1.** *Given a matroid $M = (E, \mathcal{I})$ and a collection $S$ of subsets of $E$, we say that a subcollection $S^* \subseteq S$ is $r$-representative for $S$ if the following holds: for every set $Y \subseteq E$ of size at most $r$, if there is a set $X \in S$ disjoint from $Y$ with $X \cup Y \in \mathcal{I}$, then there is a set $X^* \in S^*$ disjoint from $Y$ with $X^* \cup Y \in \mathcal{I}$.*

We will use *representative set* without specifying $r$ to mean an $(r(M) - s)$-representative set. The following result is due to Marx [28], building on Lovász [26].

**Lemma 1** ([26, 28]). *Let $M$ be a linear matroid of rank $r + s$, and let $S = \{S_1, \ldots, S_m\}$ be a collection of independent sets, each of size $s$. If $|S| > \binom{r+s}{s}$, then there is a set $S_i \in S$ such that $S \setminus \{S_i\}$ is $r$-representative for $S$. Furthermore, given a representation $A$ of $M$, we can find such a set $S_i$ in time $(m + ||A||)^{\mathcal{O}(1)}$ (note that terms polynomial in $\binom{r+s}{s}$ are bounded by $m^{\mathcal{O}(1)}$).*

**Closest cuts and gammoid rank.** Our usage of representative sets and Lemma 1 is centered around the concept of *closest sets*, defined as follows. Let $D = (V, A)$ be a digraph, and $S \subseteq V$. A set $X \subseteq V$ is *closest* to $S$ if $X$ is the unique $(S, X)$-min-cut (or, if $S$ and $X$ are not disjoint, $X \setminus S$ is the unique $(S \setminus X, X \setminus S)$-min-cut). If so, we say that $X$ is a *closest set*. For any set of vertices $X$, the *induced closest set* $C(X)$ is the unique $(S, X)$-min-cut which is closest to $S$; this is well-defined by the submodularity of cuts, and can be found in polynomial time. If $X$ is a closest set, then $C(X) = X$. Note that a closest set does not need to be a cut; there may not be any vertices except for $X$ which are separated from $S$ by $X$.

Closest sets are a natural notion, but do not seem to have a fixed name; e.g., they occur in the bipedal stage of the Multicut$(k)$ algorithm of Marx and Razgon [29]. There are also similarities between closest sets and the concept of *important separators* [27]: for any set $X$ closest to $S$ which separates some vertex $v \notin X$ from $S$, there is a corresponding important separator. However, the change of focus from separation to closeness means that the concepts behave differently.

We make the following observations connecting closest sets to the rank function of a gammoid.

**Proposition 1** (∗[1]). *Let $D = (V, A)$ be a digraph with a set of source vertices $S \subseteq V$, and let $X \subseteq V$. Let $D'$ be the result of adding a sink-only copy $x'$ for every vertex $x \in X$. The following hold.*

---
[1] Proofs of statements marked with ∗ are postponed to Section 8.

1. The set $X$ is closest to $S$ in $D$ if and only if $X + x'$ is independent in the gammoid $(D', S)$ for every $x \in X \setminus S$.
2. Let $X_B$ be a maximal independent subset of $X$. A vertex $v$ is reachable from $S$ in $D - C(X)$ if and only if $X_B + v$ is independent in the gammoid $(D', S)$.

In particular, any $(S, T)$-cut $X$ which is not a closest set, i.e., such that $X + x'$ is dependent in the gammoid $(D', S)$ for some $x \in X \setminus S$, can be replaced by a different $(S, T)$-cut $Z$ of at most the same size which does not contain $x$.

## 4 Representative sets: A polynomial kernel for Almost 2-SAT

In this section we show how to obtain polynomial kernels via representative objects. We consider a problem that we call DIGRAPH PAIR CUT, which captures the iterative compression version of ALMOST 2-SAT. We show a polynomial kernel for DIGRAPH PAIR CUT, using representative sets and a gammoid representation of graph cuts. Polynomial kernels for ALMOST 2-SAT and related problems follow via kernelization-preserving reductions.

We now study the DIGRAPH PAIR CUT problem, to provide an $\mathcal{O}^*(2^k)$ time algorithm and a randomized polynomial kernelization for it. Given a digraph $D = (V, A)$ with a source vertex $s$, we say that a pair $p = \{u, v\}$, $u, v \in V$, is *reachable* in $D$ if both $u$ and $v$ are reachable from $s$ (not necessarily via disjoint paths). The problem is then defined as follows.

---

DIGRAPH PAIR CUT **Parameter:** $k$.
**Input:** A digraph $D = (V, A)$ with source vertex $s \in V$, a set of pairs $P \subseteq \binom{V}{2}$, and an integer $k$.
**Question:** Is there a set $X \subseteq V \setminus \{s\}$ with $|X| \leq k$ such that no pair in $P$ is reachable in $D - X$?

---

Note that DIGRAPH PAIR CUT can be seen as a cut-based generalization of VERTEX COVER. Specifically, if the graph $D$ is an $n$-point star with $s$ in the center, then the input is equivalent to a VERTEX COVER instance with one edge for every pair in $P$.

**Theorem 6** (∗). *The* DIGRAPH PAIR CUT *problem can be solved in time* $\mathcal{O}^*(2^k)$.

It is easy to see that optimal solutions for DIGRAPH PAIR CUT can be assumed to be closest to the source vertex $s$. We show that there exists a representative subset $P^*$ of the pairs $P$, of size $\mathcal{O}(k^2)$, which determines whether or not any pair from $P$ is reachable from $s$ under some cut $X$ closest to $s$. Given $P^*$, and a gammoid encoding of the cut function between $s$ and the vertices occurring in $P^*$, we get a polynomial kernel for DIGRAPH PAIR CUT. Note that the bound on $|P^*|$ is tight even in the case of VERTEX COVER, where $\mathcal{O}(k^2)$ edges is optimal [10].

**Lemma 2** (∗). *Let $D = (V, A)$ be a digraph, $s \in V$, $k$ an integer, and $P \subseteq \binom{V}{2}$ a set of vertex pairs. In randomized polynomial time (with failure probability exponentially small in the input size) we can find a set of $\mathcal{O}(k^2)$ pairs $P^* \subseteq P$ (the* representative pairs*), such that for any set $X \subseteq V \setminus \{s\}$ closest to $s$ of at most $k$ vertices, the graph $D - X$ contains a reachable pair $p \in P$ if and only if it contains a reachable pair $p^* \in P^*$.*

*Proof (sketch).* Proposition 1 implies that, for any cut $X$ closest to $s$, a vertex $v$ is reachable from $s$ in $G - X$ if and only if $X + v$ is independent in the gammoid $(D, S)$, where $S$ consists of $k + 1$ copies of $s$.

By studying a matroid $M$ consisting of two disjoint copies of $(D, S)$, we can extend this fact to pairs of vertices. We use Lemma 1 to generate a set of representative pairs $P^* \subseteq P$ of size $\mathcal{O}(k^2)$ with respect to independent sets of size at most $2k$ in $M$. Now if $P$ contains a pair that extends an independent set $X'$, with $|X'| \leq 2k$, then $P^*$ contains such a pair too; the crucial sets $X'$ in $M$ are those which correspond to two copies of some closest cut $X$. This completes the proof since, by Proposition 1, reachability with respect to closest cuts is equivalent to extension of the corresponding independent set. □

We note that a generalization from pairs to $q$-tuples still holds. If applied to $q$-tuples, the lemma uses polynomial time to reduce a given set of $q$-tuples to a set of $\mathcal{O}(k^q)$ representative $q$-tuples. This implies a polynomial kernel for the $q$-ary generalization of DIGRAPH PAIR CUT, where the input contains a set of $q$-tuples $Q$, and the task is to separate at least one member of every $q$-tuple from $s$. We call this variant DIGRAPH $q$-TUPLE CUT. Again, note that a bound of $\mathcal{O}(k^q)$ is tight for $q$-tuples: it is straightforward to reduce from $q$-HITTING SET for which total size of $\mathcal{O}(k^{q-\epsilon})$ for any $\epsilon > 0$ is excluded unless NP $\subseteq$ coNP/poly [10].

Now, let $(D, s, P, k)$ be an instance of DIGRAPH PAIR CUT, and let $P^*$ be a set of representative pairs of $P$. Clearly, it is enough to find a set $X$ of size $k$ such that no pair in $P^*$ is reachable in $D-X$; we show that, in turn, the existence of such a set can be encoded into a gammoid on $s$ and $\bigcup P^*$.

**Theorem 7** (∗). *There is a randomized polynomial-time compression algorithm for* DIGRAPH PAIR CUT *which given an instance* $I = (D, s, P, k)$ *and a positive real $\epsilon$ computes a compressed representation of $I$ of size $\tilde{\mathcal{O}}(k^3(k + \log 1/\epsilon))$. The success probability is at least $1 - \epsilon$, and in the remaining cases the output encodes a positive instance.*

As a corollary, by standard arguments (cf. [5]), we get a polynomial coRP-kernelization for DIGRAPH PAIR CUT. A kernelization-preserving reduction from ALMOST 2-SAT to DIGRAPH PAIR CUT (see Section 6) now finishes the kernelization of ALMOST 2-SAT, and the first half of Theorem 1. Further reductions give polynomial kernels for, among others, VERTEX COVER ABOVE LP and RHORN-BACKDOOR DELETION SET; see Section 6.

## 5 Finding irrelevant vertices: Polynomial kernels for cut problems

In this section we extend our scope by showing how to use representative sets for the identification of irrelevant vertices in terminal cut problems. In this setting, a vertex is said to be *irrelevant* if there is at least one optimal solution which does not contain the vertex. Note that it is well possible that some irrelevant vertices are needed to build an optimal solution, but that any single one of them can be avoided.

As a warm-up result, we will consider the MULTIWAY CUT WITH DELETABLE TERMINALS($k$) problem, where the solution is allowed to contain terminals as well. For this simpler problem we are able to give a representative set characterization that covers not only all non-irrelevant vertices, but in fact contains an optimal solution. Thus a single iteration of the representative sets tool will produce a (randomized) polynomial kernel with $\mathcal{O}(k^3)$ vertices.

Next, we move on to the main focus of the section, namely multiway cuts with a bounded number of terminals or partitions. We present a randomized polynomial kernel with $\mathcal{O}(k^{s+1})$ vertices for the $s$-MULTIWAY CUT($k$) problem, where the number of terminals is bounded by the constant $s$. For this problem we need a more "traditional" irrelevant vertex approach where only a single vertex is removed in each iteration.

The approach has further consequences, in the form of the cut-covering sets of Theorems 3 and 4 – we find a small set of vertices which simultaneously contains an optimal solution for *all* min-cut questions involving $T$ (and a similar conclusion for multiway cuts). In particular, this result lets us replace the gammoid encoding of graph cuts of [25] by a "proper" kernelization rule.

In Section 7 we give two further applications, namely polynomial kernels for $s$-MULTICUT$(k)$ and $\Gamma$-FEEDBACK VERTEX SET$(k)$ for groups $\Gamma$ with at most $s$ elements.

### 5.1 Multiway Cut with deletable terminals

In this section we focus on the MULTIWAY CUT WITH DELETABLE TERMINALS$(k)$ problem (DT-MWC$(k)$), where the task is to separate a set $T$ of terminals by deleting at most $k$ vertices (including terminals). It can be easily seen to be equivalent to MULTIWAY CUT restricted to terminals of degree one. The problem is NP-hard by a simple reduction from VERTEX COVER.[2]

Let $(G, T, k)$ be an instance of DT-MWC$(k)$, with $G = (V, E)$ and $T \subseteq V$. We will use the tool of representative sets to identify a set of $\mathcal{O}(k^3)$ *representative vertices* $V^*$ such that $V^* \cup T$ contains an optimal solution. But first, using a result of Guillemot [17], we limit $|T|$ in terms of $k$.

**Lemma 3** ($*$). *Let $(G, T, k)$ be an instance of* MULTIWAY CUT WITH DELETABLE TERMINALS$(k)$. *An equivalent instance $(G', T', k')$ with $k' \leq k$ and $|T'| \leq 2k'$ can be computed in polynomial time.*

For the representative set, we form a graph $G'$ by adding to every non-terminal vertex $v$ two sink-only copies $v'$ and $v''$, and form the gammoid on $G'$ with $T$ as source set. We form the set $S$ of triples $\{v, v', v''\}$ for $v \in V \setminus T$, and let $S^*$ be a $k$-representative set of $S$. Further, we let $V^* \subseteq V$ be the set of vertices $v$ with $\{v, v', v''\} \in S^*$. We know that $|V^*| = \mathcal{O}(k^3)$; we argue (again) that there is an optimal solution $X$ such that $(X \setminus T) \subseteq V^*$.

**Lemma 4** ($*$). *Let $(G, T, k)$ be an instance of* MULTIWAY CUT WITH DELETABLE TERMINALS$(k)$, *and let $X$ be a minimum size multiway cut (of size $k$) which as a secondary criterion has a maximum size intersection with $T$. Let $V^*$ be a representative set of $V$, as constructed above. Then for every $x \in (X \setminus T)$, the set $X + x' + x''$ is linked to $T$, and $X \subseteq T \cup V^*$.*

*Proof (sketch).* The proof is given by matching arguments on an auxiliary bipartite graph $H = (X' \cup T', E_H)$ where $X' = X \setminus T$ and $T' = T \setminus X$ (skipping deleted terminals $t \in T \cap X$), and with an edge $\{x, t\}$ if $x$ can reach $t$ in $G - (X - x)$. There are two crucial facts. First, any pseudomatching $M \subseteq E_H$ such that edges in $M$ share no endpoint in $T'$ corresponds to $|M|$ paths from $T'$ to $X'$ which overlap only on vertices in $X'$; this follows from the fact that each component of $G - X$ contains at most one terminal. Second, it follows from the maximum overlap of $X$ with $T$ that any subset $S \subseteq X$ has at least $|S| + 2$ neighbors in $H$, since instead of $S$ we could delete all but one terminal reachable from $S$. An application of Hall's Theorem completes the proof. □

By this lemma, we may set all vertices of $V \setminus (V^* \cup T)$ as undeletable, without changing the existence of a solution of size at most $k$. The easiest way to achieve this is by adding shortcut edges $\{u, v\}$ between any two vertices $u, v \in V^* \cup T$ which are connected by a path with internal vertices from $V'$, and subsequently deleting $V'$ from the graph. Thus we get a polynomial kernel.

**Theorem 8.** MULTIWAY CUT WITH DELETABLE TERMINALS$(k)$ *has a randomized kernel of $\mathcal{O}(k^3)$ vertices. The error probability can be made exponentially small in $n$; all errors are false negatives.*

---

[2] Given a graph $G$, create $G'$ by attaching a terminal $v'$ to each vertex $v$. Multiway cuts in $G'$ correspond to vertex covers of $G$ and vice versa.

## 5.2 Bounded terminals Multiway Cut

We will now focus on $s$-Multiway Cut($k$), i.e., the variant of Multiway Cut($k$) where the number of terminals is bounded by some fixed constant $s$; we show a randomized polynomial kernel with $\mathcal{O}(k^{s+1})$ vertices.

Unlike for Multiway Cut with deletable terminals($k$) we are not able to directly form a set of representative vertices which is guaranteed to contain an optimal solution. Instead we show how to identify irrelevant vertices via representative sets techniques, and get the kernel by iterating computation of representative vertices and deletion of a single irrelevant vertex.

We begin by identifying a condition under which at most $\mathcal{O}(k^{s+1})$ vertices are representative, and such that for any single non-representative vertex $v$ of $G$ there is an optimal solution not containing $v$. Let $T = \{t_1, \ldots, t_s\}$ be the set of terminals, arbitrarily ordered. Assume that the LP-based reductions of Guillemot [17] have been applied, so that the terminals have disjoint neighborhoods and $|N(T)| \leq 2k$ (see Lemma 3). Our condition of representativeness, called *high reachability*, is defined as follows.

**Definition 2.** *Let $(G, T, k)$ be an instance of $s$-Multiway Cut($k$) with $G = (V, E)$ and a set $T \subseteq V$ of at most $s$ terminals. Let $X \subseteq V \setminus T$ be a multiway cut of at most $k$ vertices. Furthermore, let $G'$ be the directed graph obtained from $G$ by adding one sink-only copy $v'$ for each $v \in V \setminus T$. We say that $v \in X$ is highly reachable under $X$ if $X + v' + N(t)$ is independent for every $t \in T$, in the gammoid $(G', N(T))$. We say that $v \in V$ is highly reachable if there exists some multiway cut $X$ of at most $k$ vertices such that $v$ is highly reachable under $X$.*

Intuitively, after removing any one terminal from $T$, the solution $X$ should still be independent with two vertex-disjoint paths to $v$. Though there are other ways to express this, the exact wording of Definition 2 will be useful for the group feedback vertex set problems in Section 7.2.

We may identify the highly reachable vertices in polynomial time.

**Lemma 5.** *We can identify a set of $\mathcal{O}(k^{s+1})$ vertices, in randomized polynomial time with error probability $\mathcal{O}(2^{-n})$, which contains all highly reachable vertices.*

*Proof.* Create a gammoid which is the disjoint union of $s+1$ layers $0, 1, \ldots, s$, where layers $1, \ldots, s$ are copies of the gammoid $(G', N(T))$, where $G'$ is as in Definition 2. For layer 0 we take the uniform matroid of rank $k$ on base set $V$, i.e., we take $(V, \binom{V}{\leq k})$. Then, create a set of tuples $(v(0), v'(1), v'(2), \ldots, v'(s))$ for every $v \in V(G)$, where $v(0)$ is the copy of $v$ in layer 0 and each $v'(i)$, with $1 \leq i \leq s$ is the sink-only copy of $v$ in the $i$:th matroid layer. Use Lemma 1 to identify a $\mathcal{O}(k^{s+1})$-sized set of representative tuples (note that the rank can be bounded by $\mathcal{O}(ks)$).

We show that the tuples $(v(0), v'(1), v'(2), \ldots, v'(s))$ corresponding to highly reachable vertices $v$ (with respect to some solution $X$) will be the unique choice for extending some particular independent sets. Hence they must be contained in the representative subset of the tuples.

Let $X$ be a solution to the instance, and let $v \in X$ be highly reachable under $X$. Consider the set $X^*$ which contains the following elements in the $s+1$ layers:

1. In layer 0 it contains all vertices $u \in X \setminus \{v\}$; these are at most $k-1$.
2. In each other layer $i \in \{1, \ldots, s\}$ it contains the vertices $X + N(t_i)$.

By assumption the tuple $(v(0), v'(1), v'(2), \ldots, v'(s))$ extends this set (to an independent set). Note that $v(0)$ is not among the vertices chosen for layer 0, since those are only the copies of $X \setminus \{v\}$.

Now assume that there is some vertex $u \neq v$ such that $(u(0), u'(1), u'(2), \ldots, u'(s))$ extends $X^*$. If $u \in X$, then its tuple intersects $X \setminus \{v\}$ in layer 0. If $u \notin X$, then it is reachable from only one terminal, say $t_i$, in $G - X$. Hence, in layer $i$, the vertex $u'(i)$ cannot be added to $X^*$ since we already request paths which saturate (the layer $i$ copies of) $X$ and $N(t_i)$.

Thus $v$ is the unique vertex whose tuple can extend the independent set $X^*$, implying that its tuple will be among the representative tuples computed via Lemma 2. Thus the corresponding $\mathcal{O}(k^{s+1})$ vertices contain all highly reachable vertices. □

Now, we need to show that any single vertex which is not highly reachable can be removed without harm, i.e., that such a vertex is irrelevant. It suffices to show that any optimal solution $X$ containing $v$ can be converted into a solution of the same size and avoiding $v$.

**Lemma 6** (∗). *Let $v$ be a vertex which is not highly reachable under any multiway cut of size at most $k$ and is not contained in $N(T)$. Then there is an optimal multiway cut which does not contain $v$.*

*Proof (sketch).* Let $X$ be an optimal multiway cut for $G$ with $v \in X$. Let $t \in T$ such that $X + v' + N(t)$ is not independent in $(G', N(T))$, according to Definition 2. It follows that there is a minimal cut $C$ of size at most $|X| + |N(t)|$ separating $N(T)$ and $X + v' + N(t)$. The proof goes by constructing a new multiway cut $X'$ from $C$ and $X$, such that $|X'| \leq |X|$ and $v \notin X'$. □

To conclude our result, we only need one simple reduction rule.

**Reduction Rule 1.** *Let $(G = (V, E), T, k)$ be a multiway cut instance, and let $v \in V \setminus (T \cup N(T))$ be a vertex which is not identified as potentially highly reachable by Lemma 5. Create $G'$ from $G$ by replacing $N(v)$ by a clique and removing $v$ from the graph. Return $(G', T, k)$.*

**Theorem 9** (∗). *$s$-MULTIWAY CUT$(k)$ has a randomized kernel of $\mathcal{O}(k^{s+1})$ vertices. The error probability can be made exponentially small in $n$, and errors are limited to false negatives.*

The kernel is, arguably, *near-combinatorial*, in the sense that it consists of a reduction rule which performs simple direct modifications to the input graph, while the condition of applicability for the rule may be seen as non-combinatorial, as it at the moment requires a representation of a gammoid and the multilinear algebra of Lemma 1.

### 5.3 Covering Graph Cuts

In this section, we will adapt the above approach to prove the more general Theorems 3 and 4 about covering terminal min-cuts and minimum multiway cuts. In essence, we find that the representative sets condition of Lemma 5 gives us much more power than we need for kernelization. We begin with the statement for cuts in directed graphs.

*Proof of Theorem 3.* Let $G, S, T$ be as given; we may assume w.l.o.g. that $S$ are sources and $T$ sinks in the graph (by adding source vertices before $S$ and sink vertices after $T$; this does not modify any cuts). Let a vertex $v \in V$ be *essential* if there are sets $A \subseteq S$, $B \subseteq T$ such that every minimum $(A, B)$-vertex cut contains $v$, and *irrelevant* otherwise. Let $A \subseteq S$ and $B \subseteq T$, and let $C_A$ resp. $C_B$ be the minimum $(A, B)$-vertex cut closest to $A$ resp. to $B$. The idea of the proof is that vertices essential for an $(A, B)$-cut are exactly those in $C_A \cap C_B$, and, in turn, by Proposition 1, we can use representative sets to find such vertices.

**Claim 1** (∗). *Let $r$ denote the size of a minimum $(S,T)$-vertex cut (which may intersect $S$ and $T$). We can find a set $Z \subseteq V$ of $\mathcal{O}(|S| \cdot |T| \cdot r)$ vertices which includes all essential vertices.*

Now, any vertex in $V \setminus (Z \cup S \cup T)$ may safely be made undeletable (as in Reduction Rule 1) without changing the size of any minimum $(A,B)$-vertex cut. By induction, this gives a set of $\mathcal{O}(|S| \cdot |T| \cdot r)$ vertices which covers a minimum $(A,B)$-vertex cut in $G$ for every $A \subseteq S$, $B \subseteq T$, as requested. □

By a construction in [25], we can transfer this result to preserving minimum cuts through a set of terminals, allowing for deleting terminals.

**Corollary 1** (∗). *Let $G = (V, E)$ be a directed graph, and $X \subseteq V$ a set of terminals. We can identify, in polynomial time, a set $Z$ of $\mathcal{O}(|X|^3)$ vertices such that for any $S, T, R \subseteq X$, a minimum $(S,T)$-vertex cut in $G - R$ is contained in $Z$.*

Regarding tightness, it is easy to show that $\Omega(|S| \cdot |T|)$ vertices are necessary, even in a very simple setup: Let $S$ and $T$ be disjoint sets of vertices of weight 2, and for every $u \in S, w \in T$ create a connecting vertex $v_{u,w}$ of weight 1, with $N(v_{u,w}) = \{u, w\}$. Then, the sets $A = \{u\}$ and $B = \{w\}$ show that we must include $v_{u,w}$ to preserve the unique minimum $(A,B)$-cut. This is easily converted into a setting without weights, by copying vertices in $S$ and $T$ into pairs of twins. We do not know whether the further factor of $r$ in our upper bound is necessary.

The above gives us *direct* polynomial kernels for a number of problems (as opposed to the implicit polynomial kernels resulting from polynomial compression due to NP-hardness reductions, e.g. as in [25]).

**Corollary 2** (∗). DIGRAPH PAIR CUT *has a kernel of $\mathcal{O}(k^4)$ vertices;* ALMOST 2-SAT *has a kernel of $\mathcal{O}(k^6)$ vertices. The* signed graph *generalizations of* ODD CYCLE TRANSVERSAL *and* EDGE BIPARTIZATION*, where edges are marked as even or odd and the goal is to kill all odd-parity cycles, have kernels with $\mathcal{O}(k^{4.5})$ respectively $\tilde{\mathcal{O}}(k^3)$ vertices. All the above kernels are randomized, and can be derandomized with polynomial advice.*

**Covering multiway cuts.** We now turn to a multiway cut variant of the above. To state our results, we define a multiway cut of a partition as follows: Let $\mathcal{X} = (X_1, \ldots, X_s)$ be a tuple of pairwise disjoint sets of vertices, $X := \bigcup X_i$. A *multiway cut of $\mathcal{X}$* is a set of vertices $C \subseteq V$, which may intersect $X$, such that for $i \neq j$, there is no path between $(X_i \setminus C)$ and $(X_j \setminus C)$ in $G - C$. In other words, it is a multiway cut of the instance produced by adding terminals $t_1, \ldots, t_s$, with $N(t_i) = X_i$. We show that we can find a set which covers a minimum multiway cut for *every* partition $\mathcal{X} = (X_1, \ldots, X_s)$ with $\bigcup_i X_i \subseteq X$.

*Proof of Theorem 4.* The proof will again be an irrelevant vertex construction, using the representative set condition of Lemma 5 as guide. Call a vertex $v \in (V \setminus X)$ *essential* if there is a partition $\mathcal{X} = (X_1, \ldots, X_s)$ of $X' \subseteq X$ such that every minimum multiway cut of $\mathcal{X}$ contains $v$. As in Section 5.2, we find that such an essential vertex must be highly reachable under the partition $\mathcal{X}$. Essentially, what remains to be proven is that Lemma 5 does not really depend on knowing the partition $\mathcal{X}$ in advance.

**Claim 2** (∗). *We can identify (in polynomial time) a set of $\mathcal{O}(|X|^{s+1})$ vertices which includes all essential vertices.*

The rest of the proof now proceeds as for Theorem 3. □

# 6 Implications for Almost 2-SAT and related problems

## 6.1 Almost 2-SAT Kernelization

We will now give a randomized polynomial kernel for ALMOST 2-SAT, by reducing the iterative compression form of the problem to DIGRAPH PAIR CUT. By bootstrapping the compression step with an approximate solution of appropriate size, we get a polynomial kernel.

Let us begin with explicitly defining the iterative compression form of ALMOST 2-SAT. For a 2-CNF formula $\mathcal{F}$, let a *deletion set* $X$ be a set of variables of $\mathcal{F}$ such that removing every clause incident to $X$ leaves a satisfiable instance. Note that the problem may alternatively be defined by deleting at most $k$ *clauses*; these two forms are equivalent under simple transformations (cf. [29]).

---
ALMOST 2-SAT COMPRESSION **Parameter:** $|X|$.
**Input:** A 2-CNF formula $\mathcal{F}$, an integer $k$, and a deletion set $X$ for $\mathcal{F}$.
**Question:** Is there a deletion set $X'$ for $\mathcal{F}$ of size at most $k$?

---

We now show the reduction.

**Lemma 7.** *There is a polynomial parameter transformation from* ALMOST 2-SAT COMPRESSION, *with input solution $X$ and parameter $k$, to* DIGRAPH PAIR CUT *with parameter $k' = |X| + k$.*

*Proof.* Let $\mathcal{F}$ be a 2-CNF formula and $X$ a deletion set for $\mathcal{F}$; assume that $|X| > k$ or else return a dummy **yes**-instance (this ensures a polynomially bounded parameter growth). Further, assume (by variable negation) that the remaining formula after $X$ has been deleted is zero-valid, i.e., satisfied by the assignment which sets all variables to false. Let $V$ be the set of variables of $\mathcal{F}$. We define a DIGRAPH PAIR CUT instance $(D, s, P, k)$ as follows. Let the vertex set of $D$ be $(V \setminus X) \cup \{s\} \cup \{x_0, x_1 : x \in X\}$. Initialize $P$ to $\{\{x_0, x_1\} : x \in X\}$ and create edges $(s, x_i)$ for $i = 0, 1$ for every $x \in X$. Thus, variables $x \in X$ are split into literals, and we require that at least one of the literals for every $x \in X$ is removed, implying either an assignment to $x$ or that $x$ is deleted (where the latter requires two vertex deletions). By the assumption of zero-validity, we are now able to use directed edges and vertex pairs to enforce that any solution to the DIGRAPH PAIR CUT instance, removing $|X| + t$ vertices for $t \leq k$, corresponds to a deletion set of size $t$ for $\mathcal{F}$.

Concretely, transfer $\mathcal{F}$ to the DIGRAPH PAIR CUT instance as follows. For every 2-clause in $\mathcal{F}$ contained in $V \setminus X$, transfer it unchanged: put implications as directed edges in $D$, and negative 2-clauses as pairs in $P$. Note that positive 2-clauses do not occur here by being zero-valid. For clauses with one endpoint in $X$, create edges and pairs depending on the type of clause: assume in the following that $x \in X$ and $y \notin X$. For a clause $(x \to y)$, create a directed edge $(x_1, y)$; for a clause $(y \to x)$, create a pair $\{x_0, y\}$; for a clause $(x \lor y)$, create a directed edge $(x_0, y)$; and for a clause $(\neg x \lor \neg y)$, create a pair $\{x_1, y\}$. Finally, for 2-clauses within $X$, create pairs, e.g., a clause $(x \to x')$ becomes a pair $\{x_1, x'_0\}$. Note that each edge incident on a variable $x_i$ in $D$ is either $(s, x_i)$ or an outgoing edge $(x_i, y)$. We argue that the resulting DIGRAPH PAIR CUT instance has a solution of size $k + |X|$ if and only if $\mathcal{F}$ has a deletion set of size $k$.

On the one hand, let $X'$ be a deletion set for $\mathcal{F}$, and let $f$ be a satisfying assignment to the remaining formula. Create a set of vertices $Z$ in $D$ as follows: for every $x \in X \cap X'$, add $x_0$ and $x_1$ to $Z$; for $x \in X \setminus X'$, with $f(x) = i$, add $x_{1-i}$ to $Z$; and for every $y \in X' \setminus X$, add $y$ to $Z$. Now $Z$ has the right size. Assume that $Z$ is not a valid solution to the DIGRAPH PAIR CUT instance, i.e., that some pair $p \in P$ is reachable in $D - Z$. First, we find that every vertex in $V \setminus X$ which is reachable in $D - Z$ must be true in $f$, as otherwise $f$ would falsify a clause. Now consider the pair $p$. In every case, there is a corresponding clause in $\mathcal{F}$ which is falsified by $f$.

On the other hand, let $Z$ be a solution to the DIGRAPH PAIR CUT instance, and let $X^*$ contain those variables $x \in X$ for which both literals are deleted. Create an assignment $f$ to variables corresponding to non-deleted vertices, by letting $f(x) = i$ if $x_i$ is non-deleted, and for $y \in V \setminus X$, let $f(y) = 1$ if $y$ is reachable from $s$ in $D - Z$, and $f(y) = 0$ otherwise. Let the deletion set be $X' = (Z \cap (V \setminus X)) \cup X^*$. Now $f$ assigns a unique value to every variable not in $X'$; consider a 2-clause in $\mathcal{F}$ falsified by $f$. Every such clause corresponds to a directed edge or a pair, in a way which would contradict $Z$ being a solution to the DIGRAPH PAIR CUT instance.

Thus we find that the reduction is correct. □

By this reduction and Theorem 7, we get a randomized polynomial-time compression of ALMOST 2-SAT COMPRESSION to size $\tilde{\mathcal{O}}(|X|^4)$. To finalize the ALMOST 2-SAT kernel, we only need an initial deletion set $X$ of size poly($k$) to bootstrap the kernelization from. By Agarwal et al. [1], Almost 2-SAT has an $\mathcal{O}(\sqrt{\log n})$-approximation (recall that the clause- and variable-deletion forms are equivalent). Using the same tricks as for ODD CYCLE TRANSVERSAL [25], we may combine this with an FPT algorithm to get a solution of size $\mathcal{O}(k^{1.5})$. We may then get a polynomial kernel in the form of an ALMOST 2-SAT instance via an NP-completeness back-reduction.

**Theorem 10.** *Almost 2-SAT has a randomized polynomial-time compression to size $\tilde{\mathcal{O}}(k^6)$, implying a polynomial coRP-kernelization.*

## 6.2 Implications of the Kernel

In this section, we show some consequences of the ALMOST 2-SAT kernel. As ALMOST 2-SAT is quite a general problem, several interesting problems have been shown to reduce to it via parameter-preserving reductions. Many of these reductions are surveyed in [38]. We get the following corollary.

**Corollary 3.** *The following problems have randomized polynomial kernels:* VERTEX COVER ABOVE MATCHING, VERTEX COVER PARAMETERIZED BY KÖNIG DELETION SET, KÖNIG VERTEX DELETION *restricted to input graphs having perfect matchings, and* RHORN-BACKDOOR DELETION SET.

*Proof.* All listed problems are NP-hard even with parameter value encoded in unary (since the maximum cost of any solution is already polynomially bounded). It suffices to refer to the appropriate polynomial parameter transformations (cf. [5]): VERTEX COVER ABOVE MATCHING and ALMOST 2-SAT($k$) are equivalent under PPTs [38]. The following problems are reducible to VERTEX COVER ABOVE MATCHING under PPTs: KÖNIG VERTEX DELETION restricted to input graphs having perfect matchings [31] and RHORN-BACKDOOR DELETION SET [16].

For VERTEX COVER PARAMETERIZED BY KÖNIG DELETION SET, i.e., given $X$ such that $G - X$ is a König graph, parameterized by $|X|$, one may observe the following: If the target value, say $\ell$, is larger than a minimum vertex cover of $G - X$ (equal its maximum matching size, say $m(G - X)$) plus $|X|$ then the answer is trivially **yes**. Otherwise, from $\ell < |X| + m(G - X)$ it follows that $\ell - m(G) \leq \ell - m(G - X) < |X|$. Hence, by accounting for the target value as $k := \ell - m(G)$, i.e., "above maximum matching", the parameter can only decrease and we have an immediate PPT. □

The result for VERTEX COVER PARAMETERIZED BY KÖNIG DELETION SET generalizes the best known kernelization result for VERTEX COVER PARAMETERIZED BY A FEEDBACK VERTEX SET [22]. There the input contains $X$ such that $G - X$ is a forest, parameterized by $|X|$; as forests are also König, the parameter can only decrease by reducing to parameterization by König deletion set. The same is true in comparison to parameterization by an odd cycle transversal $X$.

We also provide a new reduction result in the form of a PPT from VERTEX COVER ABOVE LP to VERTEX COVER ABOVE MAXIMUM MATCHING. This implies that the two problems are equivalent; the reverse direction holds trivially since the parameter can only decrease when comparing the target value to the LP cost instead of to a maximum matching.

**Lemma 8.** *There is a polynomial parameter transformation from* VERTEX COVER ABOVE LP *to* VERTEX COVER ABOVE MAXIMUM MATCHING.

*Proof.* Let $(G, k)$ be an instance of VERTEX COVER ABOVE LP, i.e., asking whether $G$ has a vertex cover of size at most $LP(G) + k$, where $LP(G)$ denotes the minimum cost of a fractional vertex cover for $G$. Let $\ell = LP(G) + k$. It is well known that the linear programming relaxation for VERTEX COVER is half-integral, so we will only consider such solutions.

It suffices for us to show that $G$ contains a large matching. For the argument let us fix any minimum integral vertex cover $X$ of $G$, and assume that the instance is **yes**, so $|X| \leq \ell$. Create a graph $G'$ by adding a set of $2k+1$ vertices $Y$, and making $X$ and $Y$ fully adjacent. Let us consider optimal fractional vertex covers of $G'$. We may clearly assume that each vertex of $Y$ is assigned the same value in such a cover. We also note that $X$ is a vertex cover of $G'$, hence $LP(G) \leq |X|$.

If all vertices of $Y$ are assigned $\frac{1}{2}$, or all are assigned 1, then the total cost is at least

$$LP(G) + \tfrac{1}{2}|Y| > LP(G) + k \geq \ell \geq |X|,$$

using the simple fact that $G'$ always incurs a cost of at least $LP(G)$ on the induced subgraph $G$ in $G'$. Thus, optimal fractional covers of $G'$ will assign zero to all of $Y$, since otherwise a cost exceeding $X$ is incurred. Setting $Y$ to zero already forces setting $X$ to one, giving the unique optimal fractional solution (with all further vertices also set to zero).

Thus the graph $G'$ has a vertex cover $X$ of size matching the minimum fractional cost $LP(G')$. It is easy to see that it also has a matching of size $|X| = LP(G')$: Assume that the vertices of $X$ cannot be matched into $V(G') \setminus X$. It follows, by Hall's Theorem, that there is a subset $X'$ of $X$ with $|N_{G'}(X')| < |X'|$. But then assigning $\frac{1}{2}$ to $X'$ and $N(X')$ would give a cheaper fractional solution, a contradiction.

It follows that $G'$ has a matching of size $|X|$, implying that $G = G' - Y$ has a matching $M$ of size at least $|X| - |Y| \geq LP(G) - (2k+1)$. Thus we get that our target value $\ell$ exceeds the size of a maximum matching by at most $\ell - |M| \leq \ell - (LP(G) - (2k+1)) = \ell - LP(G) + 2k + 1 \leq 3k + 1$.

Note that we do not need to know whether $(G, k)$ is **yes**, nor do we need to know $X$. We may compute a maximum matching $M$ of $G$ and check whether it meets at least the bound which we proved for the **yes**-case. If so, then $(G, k')$ with $k' = \ell - |M| = LP(G) + k - |M|$ is an equivalent instance of VERTEX COVER ABOVE MAXIMUM MATCHING with $k' \leq 3k + 1$. Otherwise, we know that $(G, k, \ell)$ is **no**, and we return a dummy **no**-instance. □

As an immediate corollary we get that VERTEX COVER ABOVE LP admits a polynomial kernel.

**Corollary 4.** VERTEX COVER ABOVE LP *admits a randomized polynomial kernelization.*

## 7 Implications for multi-terminal cut problems

### 7.1 Bounded terminal pairs multicut

Now, we can extend the result for $s$-MULTIWAY CUT$(k)$ to also give a polynomial kernel for $s$-MULTICUT$(k)$, i.e., MULTICUT$(k)$ restricted to instances having at most $s$ terminal pairs. The key fact is that an optimal multicut $X$ for some instance $(G, T, k)$, where $T \subseteq \binom{V}{2}$ of size at most $s$, is also a multiwaycut for some partition of the terminal vertices $V(T) = \{v \mid \exists u : \{u, v\} \in T\}$ in which no set contains both vertices of some terminal pair. By considering partitions that are maximally coarse, i.e., merging any two sets would violate this condition, we get a relation to multiway cut instances with up to (roughly) $\sqrt{2s}$ terminals.

**Theorem 11.** *$s$-MULTICUT$(k)$ admits a randomized polynomial kernel with $f(s)k^{\lceil\sqrt{2s}\rceil+1}$ vertices. The error probability can be made exponentially small in $n$, and errors are limited to false negatives.*

*Proof.* Let $(G, T, k)$ be an instance of $s$-MULTICUT$(k)$ with $T = \{\{a_1, b_1\}, \ldots, \{a_r, b_r\}\}$, for $r \leq s$. Consider an optimal multicut $X$ of size at most $k$. Clearly, in $G - X$ there is no path from $a_i$ to $b_i$ for any $i$. Let $\mathcal{T} = T_1 \cup \ldots \cup T_p$ be any partition of $\{a_1, \ldots, a_r, b_1, \ldots, b_r\}$ into $p$ nonempty sets, such that there is no path from $u$ to $v$ in $G - X$ for any $u \in T_i$ and $v \in T_j$ with $i \neq j$. Hence $X$ is a multiway cut of $\mathcal{T} = (T_1, \ldots, T_p)$. Observe also that merging any two sets $T_i$ and $T_j$ gives a coarser partition, for which $X$ is still a multiway cut.

We will now make $\mathcal{T}$ maximally coarse, under the restriction that no two terminals $a_i$ and $b_i$ are contained in the same set of the partition. This is easily achieved by iteratively merging any two sets of $\mathcal{T}$ which do not (together) contain both $a_i$ and $b_i$ for any $i$. We obtain a partition $\mathcal{T}^* = (T_1', \ldots, T_{p^*}')$ such that no two sets can be merged since they together contain a terminal pair (and recall that $X$ is a multiway cut for $\mathcal{T}^*$). It follows that $p^* \leq \lceil\sqrt{2s}\rceil$, since there are $r \leq s$ terminal pairs and each pair $T_i', T_j'$ is prevented from merging by a different terminal pair (so if $p^* \geq \sqrt{2s}+1$, there would have to exist $\binom{\sqrt{2s}+1}{2} > s$ terminal pairs). Thus, $X$ is a multiway cut of $\{a_1, \ldots, a_r, b_1, \ldots, b_r\}$ into at most $\lceil\sqrt{2s}\rceil$ groups.

To complete the proof, we can apply (i) the approach used for $s$-MULTIWAY CUT$(k)$ or (ii) Theorem 4 for covering multiway cuts. We will briefly sketch both options.

(i) To use the $s$-MULTIWAY CUT$(k)$ approach we need to change the way of identifying an irrelevant vertex. Essentially, for each partition $\mathcal{T}$ of $\{a_1, \ldots, a_r, b_1, \ldots, b_r\}$ into $p \leq \lceil\sqrt{2s}\rceil$ sets, none of them containing a terminal pair, we have to preserve the possibility of having a multiway cut of size at most $k$. Thus, for each partition, identify each set into a single terminal, and generate a set of representative vertices (covering the highly reachable vertices). However, only those vertices are declared irrelevant which are not in this set for *any of the partitions*. Since there are at most $\lceil\sqrt{2s}\rceil^{2s}$ such partitions, there are at most $\mathcal{O}(\lceil\sqrt{2s}\rceil^{2s} k^{\lceil\sqrt{2s}\rceil+1})$ vertices which are contained in at least one representative set. Hence, by iterating this procedure, removing one irrelevant vertex each time (and making its neighbors a clique), we reduce to that number of vertices.

(ii) To apply Theorem 4 we need to make the vertices in $\{a_1, \ldots, a_r, b_1, \ldots, b_r\}$ effectively undeletable. To this end, we simply create a total of $k+1$ copies of each vertex and connect them into a clique. We apply the theorem to the set $\{a_1(1), \ldots, a_1(k+1), a_2(1), \ldots, b_r(k+1)\}$ (of size $2r(k+1) \leq 2s(k+1)$) for partitions into up to $\lceil\sqrt{2s}\rceil$ sets. We get a set $Z$ of size $\mathcal{O}((2s(k+1))^{\lceil\sqrt{2s}\rceil+1})$ containing a multiway cut for each such partition.

For both options, errors can only be introduced by making too many vertices effectively undeletable (for (ii) this happens when $Z$ does not contain an optimal solution), which can result in false negatives but no false positives. $\square$

## 7.2 Group feedback vertex set for fixed groups

As a further consequence we show how to obtain a polynomial kernel for the following group cut problem. Let $(\Gamma, \otimes)$ be a finite group with unit element $1_\Gamma \in \Gamma$, and let $\Gamma = \{\alpha_1, \ldots, \alpha_s\}$ where $\alpha_1 = 1_\Gamma$.

---
$\Gamma$-FEEDBACK VERTEX SET($k$) **Parameter:** $k$.
**Input:** A directed graph $D = (V, A)$, an edge labeling $\phi : A \to \Gamma$, and an integer $k$.
**Question:** Is there a set $X$ of at most $k$ vertices such that $D - X$ has a consistent labeling, i.e., a function $\pi : V \to \Gamma$ such that $\pi(u) \otimes \phi((u,v)) = \pi(v)$ for all arcs $(u,v) \in A \setminus F$.

---

Equivalently to the above, one may ask for a set of at most $k$ vertices which meets all *nonnull* cycles, i.e., cycles in the underlying undirected graph whose edge labels sum to $\alpha \neq 1_\Gamma$; the label of an edge which is used in reverse direction is thereby interpreted as the inverse. This and similar problems were, e.g., studied by Guillemot [17] and Chudnovsky et al. [6]. Note that ODD CYCLE TRANSVERSAL is an instance of this problem, with the group $Z_2$. Guillemot [17] shows that GROUP FEEDBACK ARC SET($k$) and GROUP FEEDBACK VERTEX SET($k + |\Gamma|$) (see below) are FPT, but leaves open the questions of polynomial kernels, and whether GROUP FEEDBACK VERTEX SET($k$) is FPT. Our current results make partial progress towards the kernelization questions, but note that there is an easy reduction from MULTIWAY CUT($k$) to GROUP FEEDBACK VERTEX SET($k$): Use a group of size at least matching the number of terminals. Make a clique on the terminals and add correct edge labels to enforce a different value for each terminal. (Also make the terminals "heavy" by making budget+1 copies.) All other edges have the unit label. It is easy to verify that group cuts correspond directly to multiway cuts.

Other variants of the problem include Feedback Arc Set variants, where we ask to remove a set of $k$ edges instead of vertices, and GROUP FEEDBACK VERTEX SET, where the group $\Gamma$ is a part of the input (or even given via oracle access). Note that the edge deletion variant reduces to the vertex deletion variant through standard observations.

We remark that the graph is directed only as a technical requirement, i.e., to define a direction for the element labels. In all practical aspects (when referring to cycles in $D$, etc), we will treat the graph as an undirected graph (the underlying undirected graph of $D$), with edge labels that read differently in different directions. Thus, if $(u, v) \in D$ with label $\alpha$, then we act as if in addition $(v, u) \in D$ with label $\alpha^{-1}$. Since we are dealing with the vertex deletion variant, this does not cause any problems.

As a first step towards our kernel we will require an efficient procedure for producing an initial solution of size $k^{\mathcal{O}(1)}$. This we do by reducing our problem to a VERTEX MULTICUT setting, by splitting every vertex in $D$ into $s$ copies, in a way that can be seen as splitting vertices into "literals". Known approximation results for VERTEX MULTICUT then provide us with the initial solution we need.

**Lemma 9.** *Let $(D, \phi, k)$ be an instance of $\Gamma$-FEEDBACK VERTEX SET($k$), for a fixed group $\Gamma$ with $s$ elements. There is a polynomial-time procedure that either proves that the instance is negative or produces a solution of size $\mathcal{O}(s^2 k^2)$.*

*Proof.* Let $D = (V, A)$. We construct a graph $G = (V', E)$ with $V' = \{v(\alpha) : v \in V, \alpha \in \Gamma\}$ and $E = \{\{u(\alpha), v(\alpha \otimes \phi((u,v)))\} : (u,v) \in A, \alpha \in \Gamma\}$. These edges are chosen such that if $\{u(\alpha), v(\alpha')\} \in E$, and if $\pi$ is a consistent labeling which labels both $u$ and $v$, then $\pi(u) = \alpha$ if and only if $\pi(v) = \alpha'$.

Finally, the set of terminal pairs is $T = \{(u(\alpha), u(\alpha')) : u \in V, \alpha, \alpha' \in \Gamma, \alpha \neq \alpha'\}$. We show that the optimal multicut of $(G, T)$ is within a factor of $s$ of the optimal feedback vertex set of $(D, \phi)$.

**Claim 3.** *Let $X$ be a feedback vertex set for $(D, \phi)$ of size $k$. Then $X' = \{v(\alpha) : v \in X, \alpha \in \Gamma\}$ is valid multicut of $(G, T)$ of size $sk$.*

*Proof.* Any path between terminal pairs in $G - X'$ forms a non-null cycle in $D - X$. □

**Claim 4.** *Let $X$ be a multicut of $(G, T)$. Let $V(X) = \bigcup_{v(\alpha) \in X} \{v\}$. Then $V(X)$ is a solution to the group cut problem of size at most $|X|$.*

*Proof.* Consider a non-null cycle in $D - V(X)$, and a member $v$ of this cycle. Arbitrarily select a literal $v(\alpha)$ of $v$, and let $S$ be the set of literals propagated along the cycle, i.e., if $u$ is neighbor to $v$ in the cycle with an edge that proscribes $\pi(u) = \pi(v) \otimes \alpha'$, then add the literal $u(\alpha \otimes \alpha')$ to $S$. Do this for one revolution around the cycle, so that $S$ contains two literals of $v$. The literals of $S$ form a path in $G - X$ between different literals of $v$, which contradicts $X$ being a multicut. □

Now, if $(D, \phi, k)$ is positive, then an algorithm of Gupta [19] can be used to produce a multicut $X$ for $(G, T)$ of size $\mathcal{O}(s^2 k^2)$, which can then be projected back into an initial feedback vertex set $V(X)$ for $(D, \phi)$ of the same size. Otherwise, if $X$ is too big, we may reject $(D, \phi, k)$. □

For the remainder of the section let us fix one instance $(D, \phi, k)$ of $\Gamma$-FEEDBACK VERTEX SET$(k)$ and let $X$ be an approximate solution of size at most $\mathcal{O}(s^2 k^2)$ obtained via Lemma 9. A simple normalization is possible, in the form of an *untangling*: Fix any consistent labeling $\pi \colon (V \setminus X) \to \Gamma$ for $D - X$. We would like to get an equivalent instance but in such a way that assigning $1_\Gamma$ to all vertices is feasible for $D - X$. To this end, intuitively, we add to each vertex $v$ the inverse of $\pi(v)$ and make the correct changes for all arcs incident on $v$: For arcs $(u, v)$ we add the inverse of $\pi(v)$ to $\phi((u, v))$ (thereby defining a new arc labeling $\phi'$). For arcs $(v, u)$ we add $\pi(v)$ to $\phi((v, u))$. Note that this is performed taking all of $D$ into consideration, not only $D - X$. It can be easily verified that doing this vertex by vertex does not change the existence of a consistent assignment of $D - Y$ for any set of vertices $Y$. In the end, all arcs in $D - X$ will be labeled with $1_\Gamma$ and assigning $1_\Gamma$ (or any other single group element) to all vertices is a consistent labeling of $D - X$. The labels of arcs incident to $X$ are arbitrary. For ease of notation, we will assume that this operation has already been performed on $(D, \phi, k)$ (as a nice side effect the orientation of arcs in $D - X$ becomes immaterial as $1_\Gamma$ is self-inverse). We also assume that arcs with exactly one endpoint in $X$ are outgoing from $X$.

Algorithmically, this implies that we may solve the problem by a set of calls to MULTIWAY CUT with $s$ terminals: Consider each of the $(s + 1)^{|X|}$ possible decisions to either delete $x \in X$ or fix an assignment $\pi(x)$. Then, for every neighbor $v$ of a non-deleted vertex $x \in X$, this implies an assignment to $v$. Thus, create a graph with $s$ terminals $\alpha_1, \ldots, \alpha_s$, by starting from the underlying undirected graph of $D - X$, adding said terminals, and adding an edge $(\alpha, v)$ for every vertex $v$ such that some assignment $\pi(x)$ implies $\pi(v) = \alpha$. It is not hard to see that asking for multiway cuts over the resulting graphs produces a feedback vertex set for $(D, \phi, k)$. To incorporate this into a kernel, we need some modifications to incorporate the vertices of $X$, but the idea is the same.

**Definition 3.** *Let $(D, \phi, k)$ be a $\Gamma$-FEEDBACK VERTEX SET$(k)$ instance, normalized with respect to a feedback vertex set $X$. The graph $G(D, \phi, X)$ is an undirected graph with vertex set $V' = (V \setminus X) \cup \{\alpha_1, \ldots, \alpha_s\} \cup \{x(\alpha, \alpha') : x \in X, \alpha, \alpha' \in \Gamma\}$, such that $G[V \setminus X]$ is the underlying* undirected graph of $D[V \setminus X]$, and with additional edges as follows. For every $x \in X$ and $\alpha, \alpha' \in \Gamma$, connect $x(\alpha, \alpha')$ to terminal $\alpha \otimes \alpha'$. Furthermore, for every arc $(x, v)$ in $D$ with $v \notin X$, and every $\alpha \in \Gamma$, add an edge $\{x(\alpha, \phi(x, v)), v\}$. The terminals of $G(D, \phi, x)$ are $\{\alpha_1, \ldots, \alpha_s\}$.

Let $G = G(D, \phi, X)$ and let $T$ be the set of terminals. Now feedback vertex sets in $(D, \phi)$ correspond to certain multiway cuts in $(G, T)$, in a way that we define next.

**Definition 4.** Let $(G, T)$ be as above. A multiway cut $Z$ of $(G, T)$ is *consistent* if the following hold: For every $x \in X$, there is at most one $\alpha \in \Gamma$ such that some vertex $x(\alpha, \alpha')$ in $V'$ does not occur in $Z$, and furthermore, letting $\pi(x) = \alpha$ for every $x \in X$ such that $x(\alpha, \alpha') \notin Z$ for some $\alpha' \in \Gamma$ induces a consistent partial assignment on $D[X]$. A vertex $x \in X$ is *deleted by* $Z$ if for every $\alpha \in \Gamma$ some vertex $x(\alpha, \alpha')$ occurs in $Z$. The *cost* of $Z$ equals $|Z \cap V|$ plus the number of $x \in X$ that are deleted by $Z$.

Note that we slightly relax the consistency condition by not requiring that every literal of a deleted vertex $x \in X$ occurs in $Z$, as this will be convenient for the proof. The correspondence between multiway cuts in $(G, T)$ and feedback vertex sets in $(D, \phi)$ is as follows.

**Lemma 10.** $(D, \phi)$ has a feedback vertex set of size at most $k$ if and only if $(G, T)$ has a consistent multiway cut of cost at most $k$, where cost and consistency are in the sense of Definition 4.

*Proof.* On the one hand, let $Y$ be a feedback vertex set, and let $\pi$ be a consistent assignment for $D - Y$. Let $Z = (Y \setminus X) \cup \{x(\alpha, \alpha') \in V' : x \in Y\} \cup \{x(\alpha, \alpha') \in V' : \pi(x) \neq \alpha, x \notin Y\}$. Clearly, $Z$ is consistent (if it is a multiway cut) and has cost $|Y|$. We show that $Z$ is a multiway cut. In fact, we argue that for every $u \in V \setminus X$ such that $u$ is of distance two from a terminal $\alpha$ in $G - Z$, it must hold that $\pi(u) = \alpha$, and that for every pair of neighbors $u, v \in V \setminus X$ in $G - Z$ we must have $\pi(u) = \pi(v)$. The latter holds, since all edges of $D - X$ have label $1_\Gamma$. For the former, observe that the intermediate vertex must be $x(\pi(x), \pi^{-1}(x) \otimes \alpha)$ for some $x \in X \setminus Y$. But then, there is an edge $(x, u)$ in $D$ with $\phi(x, y) = \pi^{-1}(x) \otimes \alpha$, implying by consistency that $\pi(u) = \alpha$. Thus, consistency forbids that any vertex in $V \setminus X$ is reached from more than one terminal. It is clear that no other paths between terminals are possible.

On the other hand, consider a consistent multiway cut $Z$, and let $Y$ consist of those $x \in X$ deleted by $Z$ as well as $Z \cap (V \setminus X)$. It is clear that $|Y|$ equals the cost of $Z$; we will show that $D - Y$ admits a consistent labeling. For this, we invert the previous observation: let $\pi(u) = \alpha$ for all $u \in V \setminus X$ which are reachable from terminal $\alpha$ in $G - Z$, and $\pi(x) = \alpha$ for every $x \in X$ such that vertices $x(\alpha, \alpha')$ are spared by $Z$. Finally, let $\pi(u) = 1_\Gamma$ if $u$ is reachable from no terminal in $G - Z$. We argue that this is a consistent labeling. The labeling is consistent over internal edges in $D - X - Y$, and for non-reachable vertices $u$, since all corresponding arc labels are $1_\Gamma$. Further, it is consistent over edges within $D[X \setminus Y]$ by consistency of $Z$. Finally, it is consistent over edges $(x, u)$ by design, by the same argument as before. □

Thus, we need to preserve all minimal consistent multiway cuts in $(G, T)$. We show that the potentially highly reachable vertices of $(G, T)$ identified by Lemma 5 still allow for an irrelevant vertex reduction rule with respect to consistent multiway cuts.

**Lemma 11.** Let $V^*$ be the set of potentially highly reachable vertices identified by Lemma 5 applied to $(G, T)$. Let $v \in (V \setminus X) \setminus V^*$. Then $v$ is an irrelevant vertex, i.e., there exists an optimal consistent multiway cut $Z$ with $v \notin Z$.

*Proof.* Let $Z$ with $v \in Z \cap (V \setminus X)$ be a consistent multiway cut with minimum cost, in the sense of Definition 4. Secondarily, boost $Z$ to maximum cardinality, i.e., for every $x \in X$ the number of vertices $x(\alpha, \alpha')$ in $Z$ is either $s^2$ or $s(s-1)$. Now, let $Z_V := Z \cap (V \setminus X)$ and $Z_T := Z \cap N(T) = Z \setminus Z_V$. Since vertices $x(\alpha, \alpha') \in Z$ are contained in $Z_T \subseteq N(T)$, they are effectively treated differently by the condition of high reachability. We show that $Z_V$ may be exchanged for a different cut $Z_V'$, such that $Z_T \cup Z_V'$ is a consistent multiway cut of at most the same cost, with $v \notin Z_V'$.

Let $\alpha$ be such that $Z + v' + N(\alpha)$ is not independent in $(G, N(T))$. We first argue that $Z \cup N(\alpha)$ is independent: assume not, and let $C$ be an $(N(T), Z \cup N(\alpha))$-min cut closest to $N(T)$. Let $Z_T' = Z_T \cup N(\alpha)$; since $Z_T'$ occurs on both sides of the cut, clearly $Z_T' \subseteq C$. Thus $C \setminus Z_T'$ is a min-cut $(N(T) \setminus Z_T', Z_V)$, of size smaller than $|Z_V|$. Since the cost of $Z_V$ is just its cardinality, replacing $Z_V$ by $C \setminus Z_T'$ in $Z$ yields a multiway cut of smaller cost, which furthermore will be consistent. Thus $Z \cup N(\alpha)$ is independent.

Consider now an $(N(T), Z \cup N(\alpha) \cup \{v'\})$-min cut $C$ closest to $N(T)$, and let $C_V = C \cap (V \setminus X)$. We argue that $v \notin C_V$ and that $C_V$ is a valid replacement for $Z_V$ in $Z$, i.e., that $C_V$ can fill the role of the above promised $Z_V'$. As before, $Z_T \cup N(\alpha) \subseteq C$, and by independence $|C| = |Z \cup N(\alpha)|$, so $|C_V| \leq |Z_V|$. Since $C$ cuts $v'$ from $N(T) \setminus Z_T'$ and by the minimality of $C$ and independence of $Z \cup N(\alpha)$, we thus have $v \notin C_V$. As above, it is now clear that replacing $Z_V$ by $C_V$ in $Z$ gives a consistent multiway cut not containing $v$, of no larger cost than $Z$. □

Since the rank of the gammoid is $r := s^2|X|^2$, we get a polynomial kernel with $\mathcal{O}(r^{s+1}) = \mathcal{O}(f(s)k^{2s+2})$ vertices. Note that we may also get a kernel directly in the language of $\Gamma$-FEEDBACK VERTEX SET$(k)$, since an irrelevant vertex may be made undeletable in the original setting as well.

**Theorem 12.** *Let $\Gamma$ be a fixed group with $s$ elements. $\Gamma$-FEEDBACK VERTEX SET$(k)$ admits a randomized polynomial kernel with $\mathcal{O}(k^{2s+2})$ vertices. The error probability can be made exponentially small in $n$, and errors are limited to false negatives.*

## 8 Omitted proofs

### 8.1 Omitted proofs of Section 3

**Proposition 1.** *Let $D = (V, A)$ be a digraph with a set of source vertices $S \subseteq V$, and let $X \subseteq V$. Let $D'$ be the result of adding a sink-only copy $x'$ for every vertex $x \in X$. The following hold.*

1. *The set $X$ is closest to $S$ in $D$ if and only if $X + x'$ is independent in the gammoid $(D', S)$ for every $x \in X \setminus S$.*
2. *Let $X_B$ be a maximal independent subset of $X$. A vertex $v$ is reachable from $S$ in $D - C(X)$ if and only if $X_B + v$ is independent in the gammoid $(D', S)$.*

*Proof.* For the first part, assume first that the condition holds. Let $Z$ be a vertex min-cut $(S, X)$ distinct from $X$, and let $x \in X \setminus Z$. By assumption we have $|Z| = |X|$ and $x \notin S$. However, since $X + x'$ is independent in the gammoid, we have $|X| + 1$ vertex-disjoint paths from $S$ to $X$ in $D$, of which $Z$ will miss at least one. Thus $Z$ cannot exist.

On the other hand, let $x \in X \setminus S$, and assume that $X + x'$ is dependent in the gammoid. By the cut/flow duality, since the flow from $S$ to $X + x'$ in $D'$ is $|X|$, there is an $(S, X + x')$-cut $Z$ in $D'$ of size $|X|$. We claim that $x \notin Z$: since $X$ is independent, there are vertex-disjoint paths from $S$ to $X$, which are hit exactly once each by $Z$; thus $x' \notin Z$. If $x \in Z$, then by minimality of $Z$ there is a path from $S$ to $x$ avoiding $Z - x$. However, this path can also be used as a path from $S$ to $x'$ avoiding $Z$, contradicting that $Z$ is a cut. Thus $Z$ is a cut $(S, X)$ of size $|X|$, which is closer to $S$ than $X$ is. Clearly, $Z$ is also a cut in $D$.

For the second part, first assume that $X_B + v$ is independent. The cut $C(X)$ has size $|X_B|$, and separates $S$ from $X$. Thus it will not intersect the path from $S$ to $v$. On the other hand, assume that $X_B + v$ is dependent, and consider an $(S, X + v)$-cut $Z$ of size $|X_B|$. This also serves as an $(S, X)$-min-cut, which separates $v$ from $S$. Thus $C(X)$ separates $v$ from $S$. Note that using $(D, S)$ or $(D', S)$ makes no difference here. □

## 8.2 Omitted proofs of Section 4

**Theorem 6.** *The* DIGRAPH PAIR CUT *problem can be solved in time* $\mathcal{O}^*(2^k)$.

*Proof.* The algorithm uses the simple observation that any solution to DIGRAPH PAIR CUT can be transformed to a solution of at most the same size which performs a cut closest to $S$.

Let a digraph $D = (V, A)$, a vertex $s \in V$, a set of pairs $P$, and an integer $k$ be given.

1. Initialize by creating copies $S = \{s_1, \ldots, s_{k+1}\}$ of $s$ (to make $s$ undeletable), and let $T = \emptyset$.
2. Let $C$ be the min-$(S, T)$-cut closest to $S$. If $|C| > k$, **reject** the instance. If no pair is reachable in $D - C$, **accept** the instance.
3. Otherwise, let $p = \{u, v\}$ be a pair reachable in $D - C$. Branch two ways, adding either $u$ or $v$ to $T$ and return to Step 2.

Since every recursion in Step 3 increases the min-cut size $\lambda$, at most $k$ recursion steps are taken, and the algorithm runs in time $\mathcal{O}^*(2^k)$ as promised. □

**Lemma 2.** *Let $D = (V, A)$ be a digraph, $s \in V$, $k$ an integer, and $P \subseteq \binom{V}{2}$ a set of vertex pairs. In randomized polynomial time (with failure probability exponentially small in the input size) we can find a set of $\mathcal{O}(k^2)$ pairs $P^* \subseteq P$ (the representative pairs), such that for any set $X \subseteq V \setminus \{s\}$ closest to $s$ of at most $k$ vertices, the graph $D - X$ contains a reachable pair $p \in P$ if and only if it contains a reachable pair $p^* \in P^*$.*

*Proof.* Let $D$, $s$, $k$, and $P$ be as given. Split $s$ into $k + 1$ copies; let $S$ contain these vertices. Form a matroid $M$ by taking two disjoint copies of the gammoid $(D, S)$; let $v(1)$ and $v(2)$ be the copies of a vertex $v \in V$ in the first resp. the second part of this matroid. Now, form a set $\Pi$ by taking, for every $\{u, v\} \in P$, the pair $\{u(1), v(2)\}$. We show that the set $P^*$ can be formed exactly by taking a $2k$-representative subset of $\Pi$. Since the rank of $M$ is $2k + 2$, the size of $P^*$ follows from Lemma 1. Observe that we may always assume that $X$ is a closest set (to $S$), since the reachable pairs in $D - C(X)$ are a subset of those reachable in $D - X$.

Thus, let $X$ be a closest set to $S$, and let $X' = \{x(1), x(2) : x \in X\}$ contain all copies of the vertices of $X$. We show that for any pair $p = \{u, v\} \in P$, the pair $\{u(1), v(2)\}$ extends $X'$ if and only if $p$ is reachable in $D - X$. For this, first observe that $\{u(1), v(2)\}$ extends $X'$ if and only if both $u$ and $v$ extend $X$ in the gammoid $(D, S)$, by construction. This happens if and only if $p$ is reachable in $D - X$: Note that $X$ is independent in $(D, S)$, and $C(X) = X$. By Prop. 1, we thus have that $u$ is reachable in $D - X$ if and only if $X + u$ is independent in $(D, S)$, and ditto for $v$.

Thus, let $\Pi^*$ be a $2k$-representative subset of $\Pi$, and let $P^*$ contain those pairs occurring in $\Pi^*$. Assume that $D - X$ contains a reachable pair $p = \{u, v\} \in P$. By the above, we have that $\{u(1), v(2)\} \in \Pi$ extends $X'$, and since $|X'| \leq 2k$, some pair in $\Pi^*$ extends $X'$ as well. This represents some reachable pair $p^* \in P^*$, which finishes the claim.

As for running time and randomness behavior, since $M$ is itself a gammoid, we may simply apply Theorem 5 and Lemma 1. □

**Theorem 7.** *There is a randomized polynomial-time compression algorithm for* DIGRAPH PAIR CUT *which given an instance $I = (D, s, P, k)$ and a positive real $\epsilon$ computes a compressed representation of $I$ of size $\tilde{\mathcal{O}}(k^3(k + \log 1/\epsilon))$. The success probability is at least $1 - \epsilon$, and in the remaining cases the output encodes a positive instance.*

*Proof.* We begin by stating explicitly the dependency on the encoding gammoid.

**Claim 5.** *Let $(D, s, P, k)$ be an instance of* DIGRAPH PAIR CUT*, and let $P^* \subseteq P$ be the representative pairs given by Lemma 2 when applied to $D$, $s$, $k$, and $P$. Then $(D, s, P, k)$ is positive if and only if there exists a set $T \subseteq \bigcup P^*$ intersecting every $p \in P^*$ such that the size of an $(s, T)$-cut disjoint from $s$ is at most $k$.*

*Proof.* In one direction, assume that the size of an $(s, T)$ cut is at most $k$ for some $T \subseteq \bigcup P^*$. Let $C$ be an $(s, T)$-cut that is closest to $s$. We know that no pair $p^* \in P^*$ is reachable from $s$ in $D - C$, implying, by Lemma 2, that no pair $p \in P$ is reachable from $s$ in $D - C$. Thus, $C$ is a solution for the instance $(D, s, P, k)$.

On the other hand, every solution to the instance must in particular separate $s$ from at least one member of every pair $p \in P^* \subseteq P$. Thus if no considered set $T$ produces an $(s, T)$-cut of size at most $k$, then the instance is negative. □

By this claim, we only need to represent the sizes of cuts $(s, T)$ distinct from $s$, with $T \subseteq \bigcup P^*$. Furthermore, we only need to represent whether the size of such a cut is bounded by $k$ or not. Thus, create $k + 1$ parallel copies $s_i$ of $s$, and create a gammoid with source set $S = \{s_1, \ldots, s_{k+1}\}$ and sink set $T = \bigcup P^*$. By Lemma 2 and [25, Corollary 1], a representation of this gammoid with size $\tilde{\mathcal{O}}(k^3(k + \log 1/\epsilon))$ can be found in randomized polynomial time, with the requested success probability. □

### 8.3 Omitted proofs of Section 5.1

**Lemma 3.** *Let $(G, T, k)$ be an instance of* MULTIWAY CUT WITH DELETABLE TERMINALS$(k)$*. An equivalent instance $(G', T', k')$ with $k' \leq k$ and $|T'| \leq 2k'$ can be computed in polynomial time.*

*Proof.* Create an instance $(\hat{G}, \hat{T}, k)$ of MULTIWAY CUT$(k)$, with undeletable terminals, by attaching an undeletable super-terminal $t'$ to every terminal $t \in T$, where $t'$ is only neighbor to $t$. Let $\hat{G}$ be the resulting graph and $\hat{T}$ the set of new terminals. Then the solution sets of $(G, T, k)$, interpreted with deletable terminals, and $(\hat{G}, \hat{T}, k)$, interpreted with undeletable terminals, are identical. We may now use the LP-based approach of Guillemot [17], as also used in [8], to reduce the instance to at most $2k$ terminals. Let $X$ be the support of a half-integral optimum of the LP relaxation of $(\hat{G}, \hat{T})$, as in [17, Sect. 2.2]. Let $U \subseteq V$ be the vertices reachable from $\hat{T}$ in $\hat{G} - X$. Then by [17, Prop. 1], there is an optimal solution to $(\hat{G}, \hat{T})$, and thus to $(G, T)$, disjoint from $U$. Now, contract the vertices of $U \setminus T$ into $T$ in $G$ (it is clear that this is uniquely defined, as no vertex in $U$ is reachable from two terminals in $\hat{G} - X$). Note that every terminal reaches $X$ via vertices only from $U$. We may now apply the reduction rules of [8], and delete members of $X$ with at least two neighbors in $T$ (decreasing $k$ accordingly), until what remains is at most $|X| \leq 2k$ vertices, with one neighbor in $T$ each. Terminals that become isolated are naturally dropped. □

**Lemma 4.** *Let $(G, T, k)$ be an instance of* MULTIWAY CUT WITH DELETABLE TERMINALS$(k)$*, and let $X$ be a minimum size multiway cut (of size $k$) which as a secondary criterion has a maximum size intersection with $T$. Let $V^*$ be a representative set of $V$, as constructed above. Then for every $x \in (X \setminus T)$, the set $X + x' + x''$ is linked to $T$, and $X \subseteq T \cup V^*$.*

*Proof.* We begin by showing that $X + x' + x''$ is linked to $T$. Let $T' = T \setminus X$ and $X' = X \setminus T$, and create a bipartite graph $H = (X' \cup T', E_H)$ by connecting $x \in X'$ to $t \in T'$ if and only if $x$ is reachable from $t$ in $G - (X - x)$. Assume that $|X'| > 0$, since otherwise the claim is vacuous. Thus we also have $|T'| > 2$ (since otherwise $|X \cap T|$ is not maximal – note that $T \setminus \{t\}$ is a valid solution for any $t \in T$). We make the following easy claim.

**Claim 6.** *Let $M \subseteq E_H$ be a pseudomatching in $H$, i.e, every $t \in T'$ intersects at most one edge in $M$. Then there is a corresponding set of vertex-disjoint paths from distinct terminals in $T'$ to the endpoints of $M$ in $X'$.*

*Proof.* Edges in $H$ correspond to paths in $G$ which intersect $X$ only in their end points. Since every vertex in $G - X$ is reachable from at most one terminal, we may arbitrarily take one path in $G - X + x$ for every edge in $M$. □

Thus, we need to show that for every $x \in X'$ there is a pseudomatching which saturates $X'$ and intersects $x$ three times. For this, we need another claim.

**Claim 7.** *For every $S \subseteq X'$ we have $|N(S)| > |S| + 1$ in $H$.*

*Proof.* Assume that $|N(S)| \leq |S| + 1$ for some $S \subseteq X'$. Let $S' = N(S) - t$ for some arbitrary $t \in N(S)$. We claim that replacing $S$ by $S'$ in $X$ yields a valid solution, of size at most $|X|$. Clearly, no terminal outside of $N(S)$ is affected by removing $S$ from the solution (since all paths from such terminals must go via other members of $X$, which remain in the solution). Thus, the only failure would be a path between members of $N(S)$. However, only one member of $N(S)$ remains undeleted. Thus we create a new solution, which has size at most $|X|$ and intersects more terminals than $X$, contradicting our assumption on $X$. □

Now, the conclusion follows simply by Hall's condition. Create three copies $x, x', x''$ for the arbitrary vertex $x \in X'$, and ask for a matching saturating $X' + x' + x''$. The non-existence of such a matching would imply a set $S \subseteq X'$ such that $|N(S)| < |S + x' + x'| = |S| + 2$, which by the previous claim contradicts our choice of $X$.

Thus, there are vertex-disjoint paths from $T'$ to $X' + x' + x''$, and similarly (adding trivial paths for $X \cap T$) from $T$ to $X + x' + x''$, for any arbitrary $x \in X'$.

For the final part of the conclusion, we note that $\{x, x', x''\}$ uniquely extends $X - x$: For vertices $y \in X \setminus \{x\}$ the tuple $\{y, y', y''\}$ overlaps $X - x$. All other vertices $y \in V \setminus X$ are reachable from at most one terminal in $G - X$, and from at most two terminals in $G - (X - x)$ (where the second path goes through $x$). On the other hand, we just showed that $\{x, x', x''\}$ extends $X - x$, and is thus contained in the representative set. □

### 8.4 Omitted proofs of Section 5.2

**Lemma 6.** *Let $v$ be a vertex which is not highly reachable under any multiway cut of size at most $k$ and is not contained in $N(T)$. Then there is an optimal multiway cut which does not contain $v$.*

*Proof.* Let $X$ be an optimal multiway cut for $G$ with $v \in X$. Let $t \in T$ such that $X + v' + N(t)$ is not independent in $(G', N(T))$, according to Definition 2. It follows that there is a minimal cut $C$ of size at most $|X| + |N(t)|$ separating $N(T)$ and $X + v' + N(t)$. Clearly, $C$ must contain $N(t)$, so let $X' := (X \cap N(t)) \cup (C \setminus N(t))$. Then $|X'| \leq |X|$: we have $|C| \leq |X \cup N(t)| = |X| + |N(t) \setminus X|$, and $|X'| = |C| - |N(t) \setminus X|$ (by choice). We show that (1) $X'$ is a multiway cut for $G$ and (2) $v \notin X'$.

(1) Assume otherwise and let $P$ be a path connecting two terminals $t_i, t_j$ in $G - X'$ with $t_i \neq t_j$; w.l.o.g. $t_i \neq t$. It follows that $P$ contains at least one vertex of the multiway cut $X$. We know that $C = X' \cup N(t)$ separates $t_i$ from $X$. Since $X'$ does not intersect $P$, it follows that $C$ intersects $P$ only in $N(t) \setminus X$. But then a prefix of $P$ is a path from $t_i$ to $t$ which avoids $X$, a contradiction.

(2) Assume for contradiction that $v \in X'$. By minimality of $C$ it follows that some neighbor of $v$ is reachable from $T$ in $G - C$, implying that $v'$ must be in $C$ too. From minimality of $X$ we have that $X + N(t)$ is independent, implying $|X \setminus N(t)|$ vertex-disjoint paths $P_i$ from $N(T - t)$ to $X \setminus N(t)$ avoiding $N(t)$. Thus $C \setminus N(t)$ intersects all these paths $P_i$; however, since $v'$ is a sink-only copy, it cannot intersect any of the paths, and since $v \notin N(t)$ we have $|C \setminus N(t)| > |X \setminus N(t)|$. But since $C \cap N(t) = X \cap N(t)$, this contradicts the size of $C$. □

**Theorem 9.** *$s$-MULTIWAY CUT$(k)$ has a randomized kernel of $\mathcal{O}(k^{s+1})$ vertices. The error probability can be made exponentially small in $n$, and errors are limited to false negatives.*

*Proof.* We begin with the correctness of Reduction Rule 1.

**Reduction Rule 1.** *Let $(G = (V, E), T, k)$ be a multiway cut instance, and let $v \in V \setminus (T \cup N(T))$ be a vertex which is not identified as potentially highly reachable by Lemma 5. Create $G'$ from $G$ by replacing $N(v)$ by a clique and removing $v$ from the graph. Return $(G', T, k)$.*

**Claim 8.** *The instance $(G', T, k)$ produced by Rule 1 is equivalent to $(G, T, k)$, and the rule can be performed exhaustively in polynomial time.*

*Proof.* We begin by showing correctness of the replacement. For soundness, assume that the optimal solution for $G$ has size $k$, and let $X'$ be a multiway cut for $G'$ of size at most $k - 1$. Clearly, $X' + v$ is a multiway cut for $G$. However, by minimality of the solution, there exists a path between two terminals which only intersects $X' + v$ in $v$. This path still exists in $G'$ and does not intersect $X'$, which contradicts $X'$ being a solution. For completeness, by Lemma 6 we have an optimal multiway cut $X$ for $G$ with $v \notin X$. Assume that $X$ fails to be a multiway cut in $G'$, i.e., there is a path between two terminals in $G'$ avoiding $X$. But then there is a corresponding path in $G$ which only adds the vertex $v$, and thus is still disjoint from $X$.

Finally, we can perform the rule exhaustively by $n$ repetitions of the cycle of detecting highly reachable vertices and removing one irrelevant vertex. □

By this, as long as not all vertices are identified as potentially highly reachable, we may (in polynomial time) apply a reduction rule decreasing $|V(G) \setminus (T \cup N(T))|$. By our representativeness condition, at most $\mathcal{O}(k^{s+1})$ vertices are identified as potentially highly reachable, and the check can be performed in polynomial time. At most $4k$ additional vertices remain in $T \cup N(T)$.

Randomness is only required for getting a representation of a gammoid; the failure risk can be made exponentially small in the total input length with a polynomial running time (via Theorem 5), without affecting the kernel size. One-sidedness of the error follows since applications of Reduction Rule 1 cannot make an instance easier. □

## 8.5 Omitted proofs of Section 5.3

**Theorem 3.** *Let $G = (V, E)$ be a digraph and let $S, T \subseteq V$. Let $r$ denote the size of a minimum $(S, T)$-vertex cut (which may intersect $S$ and $T$). There exists a set $Z \subseteq V$, $|Z| = \mathcal{O}(|S| \cdot |T| \cdot r)$, such that for any $A \subseteq S$ and $B \subseteq T$, it holds that $Z$ contains a minimum $(A, B)$-vertex cut. We can find such a set in randomized polynomial time with failure probability $\mathcal{O}(2^{-n})$.*

*Proof.* Let $G, S, T$ be as given; we may assume w.l.o.g. that $S$ are sources and $T$ sinks in the graph (by adding source vertices before $S$ and sink vertices after $T$; this does not modify any cuts). Let a vertex $v \in V$ be *essential* if there are sets $A \subseteq S$, $B \subseteq T$ such that every minimum $(A, B)$-vertex cut contains $v$, and *irrelevant* otherwise. Let $A \subseteq S$ and $B \subseteq T$, and let $C_A$ resp. $C_B$ be the minimum $(A, B)$-vertex cut closest to $A$ resp. to $B$. The idea of the proof is that vertices essential for an $(A, B)$-cut are exactly those in $C_A \cap C_B$, and, in turn, by Proposition 1, we can use representative sets to find such vertices.

**Claim 1.** *Let $r$ denote the size of a minimum $(S, T)$-vertex cut (which may intersect $S$ and $T$). We can find a set $Z \subseteq V$ of $\mathcal{O}(|S| \cdot |T| \cdot r)$ vertices which includes all essential vertices.*

*Proof.* We will create a matroid $M$ with three layers as follows. Let $M_0$ be the uniform matroid of rank $r$ on $V$, i.e., $M_0 = (V, \binom{V}{\leq r})$. Create a graph $G' = (V \cup V', E')$ from $G$ by adding a sink-only copy $v'$ of every $v \in V$, and let $M_1$ be the gammoid $(G', S)$. Finally, create a graph $G''$ from $G$ by first reversing every edge, then adding a sink-only copy of every vertex, and let $M_2$ be the gammoid $(G'', T)$. Let $M$ be the matroid created by a direct sum of $M_0$, $M_1$, and $M_2$; for a vertex $v \in V$, denote its copy in $M_0$ by $v(0)$, its two copies in $M_1$ by $v(1)$ and $v'(1)$, and its two copies in $M_2$ by $v(2)$ and $v'(2)$. Form a set of tuples by $\mathcal{T} = \{(v(0), v'(1), v'(2)) : v \in V \setminus (S \cup T)\}$. For a vertex $v$, let $T(v) = (v(0), v'(1), v'(2))$. Let $\mathcal{T}^* \subseteq \mathcal{T}$ be a representative set for $\mathcal{T}$ in $M$, and let $V^* = \{v \in V : T(v) \in \mathcal{T}^*\}$. We show that every essential vertex is contained in $V^* \cup S \cup T$.

For this, let $A \subseteq S$ and $B \subseteq T$, and let $C$ be a minimum $(A, B)$-vertex cut. Let $v \in V \setminus (S \cup T)$ be a vertex which is a member of every minimum $(A, B)$-vertex cut. Consider the set $\hat{C}$ consisting of the copies of $(C - v)$ in $M_0$, the copies of $(S \setminus A) \cup C$ in $M_1$, and the copies of $(T \setminus B) \cup C$ in $M_2$. By assumption, $\hat{C}$ is independent. We will show that, $T(v)$ is the unique tuple in $\mathcal{T}$ which can extend $\hat{C}$.

Let $C_A$ be the unique minimum $(A, B)$-vertex cut closest to $A$; pushing $(S \setminus A) \cup C$ to a set closest to $S$ yields $(S \setminus A) \cup C_A$. Thus, by Prop. 1 (and using that $S$ are source-vertices only), an element $u'(1)$ extends $\hat{C}$ if and only if $u'$ is reachable from $A$ in $G' - C_A$. The situation with respect to $T$ is symmetric (letting $C_B$ be the min-cut closest to $B$). Thus, if $T(u)$ extends $\hat{C}$ for some $T(u) \in \mathcal{T}$, then $u'$ is reachable both from $A$ and from $B$; since $C$ is a cut, we get $u \in C_A \cap C_B$. But this implies $u \in C$, and the only vertex $u \in C$ such that $u(0)$ extends $\hat{C}$ is $v$. Thus $T(v)$ uniquely extends $\hat{C}$, and $v$ (and by extension, every essential vertex of $V \setminus (S \cup T)$) must be included in $V^*$.

Finally, the size of $Z$ would by a naïve application of Lemma 1 be $\mathcal{O}((r + |S| + |T|)^3)$; however, an inspection of the proof of Lemma 1 given in [28] reveals that a bound of $\mathcal{O}(|S| \cdot |T| \cdot r)$ suffices, since the elements of $T(v)$ are taken from different parts of a direct sum matroid. □

Now, any vertex in $V \setminus (Z \cup S \cup T)$ may safely be made undeletable (as in Reduction Rule 1) without changing the size of any minimum $(A, B)$-vertex cut. By induction, this gives a set of $\mathcal{O}(|S| \cdot |T| \cdot r)$ vertices which covers a minimum $(A, B)$-vertex cut in $G$ for every $A \subseteq S$, $B \subseteq T$, as requested. □

**Corollary 1.** *Let $G = (V, E)$ be a directed graph, and $X \subseteq V$ a set of terminals. We can identify, in polynomial time, a set $Z$ of $\mathcal{O}(|X|^3)$ vertices such that for any $S, T, R \subseteq X$, a minimum $(S, T)$-vertex cut in $G - R$ is contained in $Z$.*

*Proof.* Create a graph $G'$ by adding a source-only copy $x^-$ for every $x \in X$. For a set $A \subseteq X$, let $A^- = \{x^- : x \in A\}$. Compute a set of vertices $Z$ as in Theorem 3, using $S = X^-$ and $T = X$. We show that the vertices of $Z$ work as needed.

For that, let $S, T, R \subseteq X$, and let $A = S^- \cup R^-$ and $B = T \cup R$. We show a correspondence between minimum $(A, B)$-cuts in $G'$ and minimum $(S, T)$-cuts in $G - R$. Let $C$ be a minimum $(S, T)$-vertex cut in $G - R$, and let $C' = C \cup R$; it is clear that $C'$ is an $(A, B)$-vertex cut in $G'$. Conversely, let $C'$ be a minimum $(A, B)$-vertex cut in $G'$. Then $C := C' \setminus (R^- \cup R)$ is a valid $(S, T)$-cut in $G - R$: Assume otherwise. Then there is a path from $S$ to $T$ avoiding $C \cup R$, which also avoids $C' \subset C \cup R \cup R^-$. By minimality of $C'$, for every $x \in R$, exactly one of $x^-$ and $x$ are contained in $C'$; thus $|C' \cap (R \cup R^-)| = |R|$.

We find that $C' \setminus (R \cup R^-)$ is a minimum $(S, T)$-vertex cut in $G - R$, for any minimum $(A, B)$-vertex cut $C'$, one of which will be covered by $Z$. This finishes our proof. □

**Corollary 2.** Digraph Pair Cut *has a kernel of $\mathcal{O}(k^4)$ vertices;* Almost 2-SAT *has a kernel of $\mathcal{O}(k^6)$ variables. The* signed graph *generalizations of* Odd Cycle Transversal *and* Edge Bipartization*, where edges are marked as even or odd and the goal is to kill all odd-parity cycles, have kernels with $\mathcal{O}(k^{4.5})$ respectively $\tilde{\mathcal{O}}(k^3)$ vertices. All the above kernels are randomized, and can be derandomized with polynomial advice.*

*Proof.* The kernels for Digraph Pair Cut and Almost 2-SAT are most immediate. For Digraph Pair Cut, first reduce the pairs to a representative set of $\mathcal{O}(k^2)$ pairs, then apply the above with $k + 1$ copies of the source vertex as set $S$, and all vertices occurring in the remaining pairs as set $T$. Since the flow is bounded by $|S|$, we get $\mathcal{O}(k^4)$ vertices. For Almost 2-SAT, we get a kernel since the Digraph Pair Cut instance resulting from the reductions is a valid instance of Almost 2-SAT.

For Odd Cycle Transversal, first consider the compression form. Given a set $X$ which is a valid solution, we first identify a set $Z$ of $\mathcal{O}(|X|^3)$ vertices which covers an optimal solution for the resulting terminal cut problem in an auxiliary graph $G'$. It follows from the proof of correctness of the Odd Cycle Transversal algorithm (see also [25]) that $Z$ covers a minimum odd cycle transversal of the original graph $G$. Thus, for every path connecting $u, w \in Z$ using only vertices $v \in (V \setminus Z)$, add an even or odd edge $uw$ according to the parity of the path. The odd cycle transversal contained in $Z$ is still valid, and we certainly do not make the problem easier by reducing down to $Z$.

For Edge Bipartization, first use the approximation algorithm of Avidor and Langberg [2], providing an approximation ratio of $\mathcal{O}(\log \text{OPT})$. Let $X$ be the returned solution. If $|X| > \max(k_0, ck \log k)$ for some constants $k_0, c$, then we know that OPT is greater than $k$ and may reject the instance (or output a dummy negative instance); otherwise $|X| = \mathcal{O}(k \log k)$. We may now transform the instance into a vertex deletion instance (by subdividing edges and turning vertices into many parallel copies), in which case $X$ becomes an odd cycle transversal for the resulting OCT instance (the size of $X$ and the parameter $k$ stay the same). Proceed as for OCT, producing a cut-covering set $Z$ with $\mathcal{O}(|X|^3) = \tilde{\mathcal{O}}(k^3)$ vertices. We may assume (by the way the reduction into an OCT instance was performed) that every vertex in $Z$ corresponds to an edge in the original instance. Let $G = (V, E)$ be the original graph, and $E_Z \subseteq E$ the set of edges chosen in $Z$; call $E_Z$ the marked edges. We assume (by correctness of the construction of $Z$) that the unmarked edges span no odd cycle. Finally, we produce a kernel with edge set $E_Z$ as follows. First, partition $V$ according to components which are connected using only unmarked edges; since we will not have to delete any such edge, each such component can be collapsed into two vertices (a "negative" and a "positive" vertex). Then, we may further identify the two vertices in each component, keeping the positive, and transferring edges $E_Z$ intersecting a negative vertex into edges on positive vertices only, modifying the sign of the edge as necessary. This finishes the kernel. □

**Theorem 4.** *Let $G = (V, E)$ be an undirected graph and $X \subseteq V$. For any $s$, there exists a set $Z \subseteq V$, $|Z| = \mathcal{O}(|X|^{s+1})$, such that for any partition $\mathcal{X} = (X_1, \ldots, X_s)$ with pairwise disjoint subsets of $X$, it holds that $Z$ contains a minimum multiway cut of $\mathcal{X}$ (i.e., a minimum cut $C$ such that no pairs of sets $X_i, X_j$ are connected to each other in $G - C$). We can find such a set in randomized polynomial time with failure probability $\mathcal{O}(2^{-n})$.*

*Proof.* The proof will again be an irrelevant vertex construction, using the representative set condition of Lemma 5 as guide. Call a vertex $v \in (V \setminus X)$ *essential* if there is a partition $\mathcal{X} = (X_1, \ldots, X_s)$ of $X' \subseteq X$ such that every minimum multiway cut of $\mathcal{X}$ contains $v$. As in Section 5.2, we find that such an essential vertex must be highly reachable under the partition $\mathcal{X}$. Essentially, what remains to be proven is that Lemma 5 does not really depend on knowing the partition $\mathcal{X}$ in advance.

**Claim 2.** *We can identify (in polynomial time) a set of $\mathcal{O}(|X|^{s+1})$ vertices which includes all essential vertices.*

*Proof.* Let $G'$ be the graph $G$ with a sink-only copy $v'$ added for every $v \in V$, and a private source vertex $x^-$ added for every $x \in X$. For a set $A \subseteq X$, let $A^- = \{x^- : x \in A\}$. We create a gammoid as in Lemma 5, as a direct sum of $s+1$ layers, where layer 0 is the uniform matroid of rank $|X|$ on $V$, and layers 1 through $s$ are copies of the gammoid $(G', X^-)$. For a vertex $v$, let $v(i)$ (resp. $v'(i)$) denote the copy of $v$ (resp. the sink-only copy of $v$) in layer $i$, and let $T(v) = (v(0), v'(1), \ldots, v'(s))$. Let $\mathcal{T} = \{T(v) : v \in V \setminus X\}$. Let $\mathcal{T}^*$ be a representative set for $\mathcal{T}$, and let $V^* = \{v \in V : T(v) \in \mathcal{T}^*\}$. We show that every essential vertex is contained in $V^* \cup X$.

Let $\mathcal{X} = (X_1, \ldots, X_s)$ be a partition of $X' \subseteq X$, and let $v \in V \setminus X$ be essential in $\mathcal{X}$. Let $X_0 := X \setminus X'$. Let $C$ be a multiway cut of $\mathcal{X}$; as was established in Section 5.2, we can saturate $C + v'$ from $(X' \setminus X_i)$ for every $i \in [s]$; note that we can saturate a set $A$ from $(X' \setminus X_i)$ if and only if we can saturate $A \cup X_i^- \cup X_0^-$ from $X^-$. For a set $A \subseteq V$, let $A(i) = \{v(i) : v \in A\}$. Consider the set

$$\hat{C} := (C - v)(0) \cup \bigcup_{i=1}^{s}(C \cup X_i^- \cup X_0^-)(i).$$

As already argued, $T(v)$ extends $\hat{C}$. On the other hand, assume that $T(u)$ extends $\hat{C}$ for $u \neq v$; by layer 0, $u \notin C$. By $\hat{C}$, $u$ must be reachable from $X' \setminus X_i$ for every $i \in [s]$ (all other sources are killed at $X^-$). But then $u$ is reachable from two parts of $\mathcal{X}$, contradicting that $C$ is a multiway cut. □

The rest of the proof now proceeds as for Theorem 3. □

# 9  Conclusion

We give powerful new techniques for polynomial kernelization, centered around applications of a lemma from matroid theory due to Lovász [26] and Marx [28]. The resulting tools significantly advance the field of kernelization, and imply polynomial kernels for a range of problems, including ALMOST 2-SAT($k$), $s$-terminal MULTIWAY CUT($k$), and MULTICUT($k$) with $s$ terminal pairs, for constant $s$, among other results. In particular, we show how the lemma can be applied to find irrelevant vertices for graph cut problems. In addition to the aforementioned kernels, this lets us find a form of cut-covering sets of small size: given a graph $G$ and terminal set $T$, we can find a set $Z$, of size polynomial in $|T|$, such that for every $A, B \subseteq T$, a minimum $(A, B)$-vertex cut is contained in $Z$. Similarly, for a constant $s$, we can find a set $Z$ of polynomial size such that for every partition of $T$ (or a subset of $T$) into at most $s$ partitions, a minimum multiway cut of the partition is contained in $Z$. We foresee further applications of these results. Similarly to in [25], our kernels are randomized; unlike in [25], they can all be made reduction rule-based. Furthermore, the failure probability can be reduced to be exponentially small in the input length, implying non-uniform polynomial-time kernelization.

Despite being randomized, we note that all our kernels are compatible with the lower bound framework of Bodlaender et al. [3] and Fortnow and Santhanam [14]; see the discussion in [25] regarding coRP-kernels. Similar arguments can be made regarding non-uniform kernels; see [14, Corollary 3.4]. Hence, concrete polynomial upper and lower bounds for the current problems is a relevant path of research (see Dell and van Melkebeek [10], Dell and Marx [9], and Hermelin and Wu [21]). Research on non-uniform kernels in general is also left as future work.

Further significant open questions include the existence of polynomial kernels for the general form of MULTIWAY CUT($k$) and MULTICUT ($s+k$), in edge- and vertex-deletion variants, and for the GROUP FEEDBACK ARC SET($k$) and GROUP FEEDBACK VERTEX SET($k$) problems with arbitrary groups. Additionally, a polynomial kernel for DIRECTED FEEDBACK VERTEX SET remains an open problem.

**Acknowledgements.** The authors are grateful to Dániel Marx for comments on an early draft of the paper, and to Saket Saurabh for an early version of [32]. The second author thanks Madhusudan Manjunath for enlightening discussions.